\documentclass[apjl,letterpaper]{emulateapj}
\usepackage{subfigure}

\slugcomment{accepted for publication in \apj}

\shorttitle{Decoding SEDs of dust obscured starburst-AGN}
\shortauthors{Han \& Han}

\begin{document}
\title{Decoding spectral energy distributions of dust-obscured starburst-AGN}

\author{
Yunkun Han\altaffilmark{1,2,3},
Zhanwen Han\altaffilmark{1,3}
}
\altaffiltext{1} {National Astronomical Observatories / Yunnan Observatory, Chinese Academy of Sciences, Beijing, 100012, China}
\altaffiltext{2} {Graduate University of Chinese Academy of Sciences, Beijing, 100049, China}
\altaffiltext{3}{Key Laboratory for the Structure and Evolution of Celestial Objects, Chinese Academy of Sciences, Kunming, 650011, China}
\email{hanyk@ynao.ac.cn}

\begin{abstract}
We present BayeSED, a general purpose tool for doing Bayesian analysis of SEDs by using whatever pre-existing model SED libraries or their linear combinations.
The artificial neural networks (ANNs), principal component analysis (PCA) and multimodal nested sampling (MultiNest) techniques are employed to allow a highly efficient sampling of posterior distribution and the calculation of Bayesian evidence.
As a demonstration, we apply this tool to a sample of hyperluminous infrared galaxies (HLIRGs).
The Bayesian evidences obtained for a pure Starburst, a pure AGN, and a linear combination of Starburst+AGN models show that the Starburst+AGN model have the highest evidence for all galaxies in this sample.
The Bayesian evidences for the three models and the estimated contributions of starburst and AGN to infrared luminosity show that HLIRGs can be classified into two groups: one dominated by starburst and the other dominated by AGN.
Other parameters and corresponding uncertainties about starburst and AGN are also estimated by using the model with the highest Bayesian evidence.
We found that the starburst region of the HLIRGs dominated by starburst tends to be more compact and has a higher fraction of OB star than that of HLIRGs dominated by AGN.
Meanwhile, the AGN torus of the HLIRGs dominated by AGN tend to be more dusty than that of HLIRGs dominated by starburst.
These results are consistent with previous researches, but need to be tested further with larger samples.
Overall, we believe that BayeSED could be a reliable and efficient tool for exploring the nature of complex systems such as dust-obscured starburst-AGN composite systems from decoding their SEDs.
\end{abstract}


\section{INTRODUCTION}
The formation and evolution of galaxies and super-massive black holes (SMBHs) are now believed to be tightly related.
Meanwhile, violent formation of stars (starburst) and growth of SMBHs (AGN) are found to be coupled and ongoing together in the most infrared-luminous galaxies, such as ultraluminous infrared galaxies (ULIRGs) and hyperluminous infrared galaxies (HLIRGs).
These galaxies represent important phases in the formation and evolution of galaxies, and ideal laboratories for studying the starburst-AGN connections.
Since both of star formation and SMBHs accretion are taking place, while a large amount of dust is distributed throughout, this kind of galaxies are very complex dust-obscured starburst-AGN composite systems.
The spectral energy distributions (SEDs) are the primary source of information for our understanding of them.
However, currently it is still very challenging to efficiently extract the basic physical properties of these galaxies from the analysis of their SEDs.

The analysis of SED, or SED fitting, tries to extract one or several physical properties of a galaxy from fitting models to the observed SED.
Nowadays, new observing facilities and large surveys allow us to obtain the full SEDs at wavelengths from the X-rays to the radio for galaxies extending from local to redshifts higher than 6.
On the other hand, as the starting point of SED fitting, a library of model SEDs needs to be built in advance.
For most galaxies, stars are the main sources of lights.
The evolutionary population synthesis (EPS) models \citep{Tinsley1972a,Searle1973a,Tinsley1978a,Larson1978a,Bruzual1983a,Fioc1997a,Leitherer1999a,Bruzual2003a,Maraston2005a,Zhang2005a,Zhang2005b,Han2007a,Conroy2009a}, which are based on the knowledge of stellar evolution such as the assumed stellar initial mass function (IMF), star formation history (SFH), stellar evolutionary tracks, and stellar libraries, are standard tools for modeling the SEDs of galaxies.

Meanwhile, the dusty interstellar medium (ISM), if presented, have important effects on the resulting SEDs of galaxies.
A fraction, or most in extreme cases such as ULIRGs and HLIRGs, of the initial radiations from stars are absorbed and reprocessed by the gas and dust in the ISM.
Gases heated by young stars produce luminous nebular emission lines, while dusts heated by stars of all ages  produce the mid-infrared (MIR) and far-infrared (FIR) emission.
A simple method is to handle the absorption of star lights and their re-emission independently, and then connect them by assuming that the total energy absorbed in the UV-optical equals to the total energy re-emitted in the MIR and FIR \citep{Devriendt1999a,daCunha2008a,Noll2009a}.
A more self-consistent treatment requires detailed radiative transfer (RT) calculations to be performed using the ray-tracing \citep{Silva1998a,Efstathiou2000a,Granato2000a,Tuffs2004a,Siebenmorgen2007a,Groves2008a}, or Monte-Carlo method \citep{Baes2003a,Dullemond2004a,Jonsson2006a,Chakrabarti2009a}.
However, these RT calculations are commonly computationally expensive.

Finally, if a powerful active galactic nucleus (AGN) is presented in the center of a galaxy, the resulting SED would be largely modified.
AGN can contribute to all wavelength regimes of the electromagnetic spectrum, with accretion-disk+corona to the X-ray-UV-optical, torus to the IR, and jet to the radio and gamma-ray in some cases.
In quasars, the AGN light dominates over the integrated galaxy light at almost any wavelength, while for AGNs with lower luminosities, the contribution of the host galaxy may dominate in many wavelengths.
The modeling of the SEDs of various components of AGN have been developed independently, and all needs a special suite of parameters.
Meanwhile, the high-energy radiations from the center AGN can also be absorbed by the ISM in the host galaxy and re-emitted in the IR.
So, if AGN is presented, the relative geometry of starburst, ISM, and AGN is important for modeling the SEDs of such dust-obscured starburst-AGN composite systems.
Furthermore, if violent starburst and AGN activities are coupled and ongoing together, it may not be reasonable to model the SEDs of such systems by a simple linear combination of models for starburst and AGN developed independently.

Overall, the SEDs of dust-obscured starburst-AGN composite systems are very complicated.
A completely self-consistent SED model for such complicated systems must be very hard to be constructed.
To make progress, parameterizations of all possible components, their relative geometry and possible physical relations are inevitable. 
Given the complexities mentioned above, it is natural to expect a large number of parameters, and many possible degeneracies  between them.
Meanwhile, since the effects of dust attenuation, line emission, and dust emission have to be taken into account, the problem of determining the physical parameters from fitting model to observations is highly nonlinear.

The SED fitting methods have been improved significantly in the last decade, which allow us to extract much more complex information imprinted in the SEDs \citep[see][for a recent review of this field]{Walcher2011a}.
Among the numerous SED fitting method, we believe the method based on Bayesian inference \citep{Benitez2000a,Kauffmann2003b,Feldmann2006a,Salim2007a,Bailer-Jones2011a}, in which multi-wavelength SEDs are fitted by firstly precomputing a library of model SEDs with varying degrees of complexity and afterwards determining the model and/or model parameters that best fit the data, is the best choice for the problem we are facing.
This method gives us detailed probability distributions of model parameters and the Bayesian evidence as a quantitative evaluation of the entire model given the data.

However, the Bayesian approach requires an intensive sampling of the posterior distribution, which is a function of all parameters, and resulted from combining all priori knowledges about the parameters of the model and the new information introduced by the observations.
So, if the model used to explain the observations is itself computationally expensive, the sampling must be very time consuming.
Furthermore, as mentioned above, the model SEDs are commonly precomputed as a library, rather than computed during the sampling.
For this reason, an interpolation method must be employed.
Since the dependences of parameters with the resulting SED are highly non-linear and the number of parameters is very large, common interpolation methods are not very suitable.
This problem can be solved more easily with artificial neural networks (ANN) \citep{Lahav1996a,Bertin1996a,Andreon2000a,Firth2003a,Collister2004a,Vanzella2004a,Carballo2008a,Auld2008a}.
An ANN can be trained to approximate a library of model SEDs with highly non-linear complexities, and allow the parameter space of the model to be explored more continuously.
On the other hand, a principal component analysis (PCA) \citep{Francis1992a,Glazebrook1998a,Wild2005a,Wild2007a,Budavari2009a} can be applied to the library of model SEDs in advance to simplify the required structure of the ANN. 

These methods have been demonstrated nicely by \cite{AsensioRamos2009a}, who apply it to a recently developed public database of clumpy dusty torus model \citep{Nenkova2008a,Nenkova2008b}.
\cite{Almeida2010a} and \cite{Silva2011a} implemented ANN into the RT code GRASIL to speedup the computation of the SED of galaxies when applied to semi-analytic models \cite[SAMs;][]{White1978a,Lacey1991a,White1991a}.
The core of these methods are actually very general and can be applied to any problem regarding fitting  precomputed libraries of model SEDs to observations.
However, these methods are commonly implemented specifically for a special problem, and not convenient to be used in other similar problems.
Inspired by these works, we have built a suite of general purpose programs to generalize these methods such that they can be integrated together to do the Bayesian analysis of SEDs by comparing whatever pre-existing model SED libraries or their linear combinations with observations.

This paper is structured as follows.
In \S \ref{sect:modelSED} we describe the PCA and ANN methods used to boost the generation of model SEDs, and our implements of them.
In \S \ref{sect:BayeSED} we present our Bayesian inference tool, BayeSED.
We begin in \S \ref{ssect:bayesian} by a general introduction to Bayesian inference methods.
In \S \ref{ssect:sampling}, we discuss the posterior sampling methods.
Then, the construction of BayeSED is presented in \S \ref{ssect:BayeSED}.
In \S \ref{sect:application}, we apply this tool to the HLIRGs sample of \cite{Ruiz2007a}.
Finally, a summary of this paper is presented in \S \ref{sect:summary}.

\section{Generation of Model SEDs}
\label{sect:modelSED}
\subsection{Principal Component Analysis of Model SED Libraries}
\label{ssect:PCA}
A SED can be described by a vector of $N$ flux densities $(f_1,f_2,\cdots,f_N)$ at wavelengths $(\lambda_1,\lambda_2,\cdots,\lambda_N)$.
However, when the flux at a given wavelength is changed, the fluxes at surrounding wavelength points are also changing in a very similar way due to the continuity of the SED.
So, the fluxes at different wavelengths are not completely independent and the actual dimension of the SED is much less than $N$.
This simple fact make it possible to apply some dimensionality reduction techniques to efficiently compress the representation of a SED.

One such technique is called principal component analysis (PCA).
It can be used to derive an optimal set of linear components, called principal components, by diagonalizing the covariance matrix of a set of SEDs to find the directions of greatest variation.
Then, the original data can be approximated by a linear combination of first $N' \ll N$ independent principal components.
It is worth noting that PCA performs a linear analysis.
So, if the dependences of parameters with corresponding SED is non-linear, the number of necessary eigenvectors is commonly larger than the number of physical parameters of the model.

We adopt an IDL package for PCA \footnote{\url{http://www.roe.ac.uk/~vw/vwpca.tar.gz}}, which gathers together several  algorithms for PCA into a single package and all with the same usage. 
The Singular Value Decomposition (SVD) rather than the Robust and Iterative algorithm provided in this package \citep{Budavari2009a} is used.
The later two algorithms, when applied to the observed SEDs, can be used to obtain eigenvectors with more clear physical meanings.
However, we apply PCA to model SEDs to reduce them in a purely mathematical sense.
Meanwhile, SVD algorithm is much faster and good enough for our purpose.

We have applied PCA to two widely used model SED libraries: SBgrid for starburst galaxies and ULIRGs \citep{Siebenmorgen2007a} and CLUMPY for AGN clumpy torus \citep{Nenkova2008a,Nenkova2008b}.
The model SEDs in the SBgrid library have 5 parameters: nuclear radius $R$, total luminosity $L_{\rm tot}$, ratio of the luminosity of OB stars with hot spots to the total luminosity $f_{\rm OB}$, the visual extinction from the edge to the center of the nucleus $A_{\rm V}$, and the dust density in the hot spots  $n_{\rm hs}$ \citep[see][for detailed explanations about these parameters]{Siebenmorgen2007a}.
On the other hand, the model SEDs in the CLUMPY library have 6 parameters about the dusty and clumpy torus: the ratio of the outer to the inner radii of the toroidal distribution $Y=R_{\rm o}/R_{\rm d}$, the optical depth of clumps $\tau_{\rm V}$, the number of clouds along a radial equatorial path  $N$, the power of the power law ($r^{-q}$) describing radial density profile $q$, the width parameter charactering the angular distribution $\sigma$, and the viewing angle measured from the torus polar axis $i$.

However, it is very necessary to do some normalizations to the libraries before doing the PCA of them.
Firstly, we find that the PCA will have better performance if we use the logarithm of flux.
Secondly, the mean spectra of a SED library is found and removed from every SED in the library in advance.
These normalizations will impact on the resulting eigenvectors.
The different physical mechanisms associated with each eigenvector is not important for us, and we have not tried to find them out.
We only treat the PCA eigenvectors as a set of purely mathematical basis that allows us to efficiently reconstruct all SEDs in a library.

\begin{figure*}
  \centering 
  \includegraphics[scale=0.5]{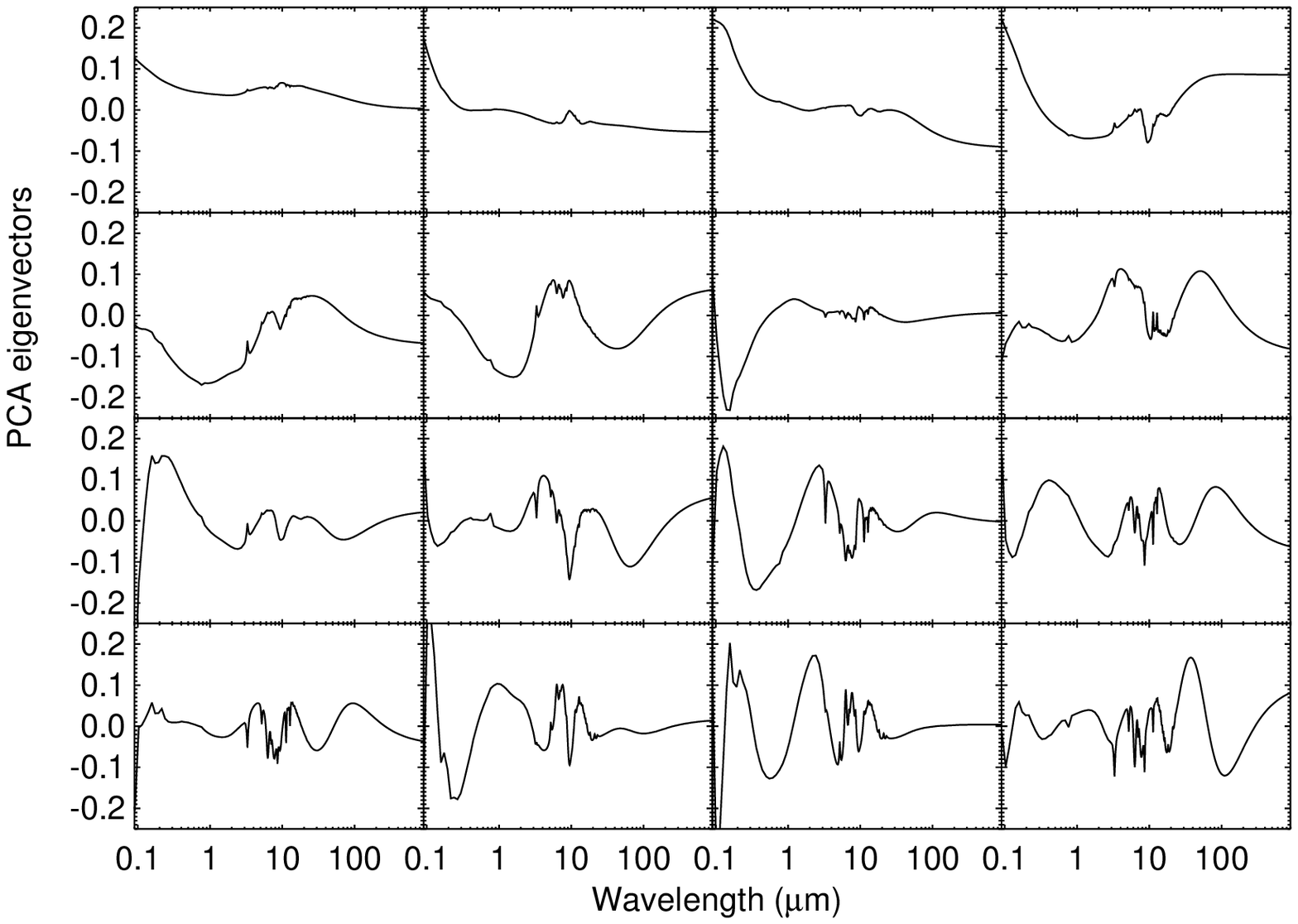}
  \includegraphics[scale=0.5]{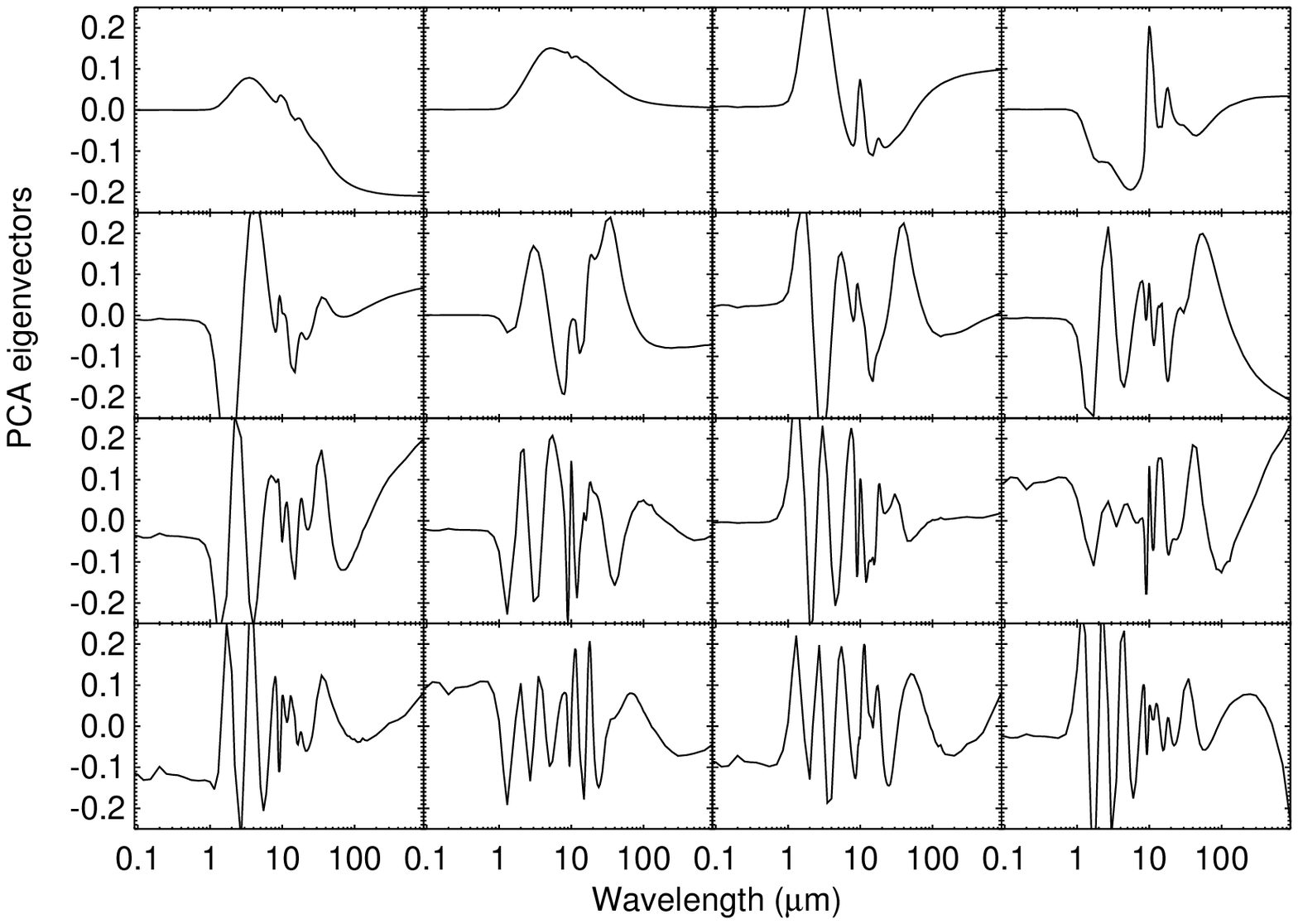}
  \caption{First 16 eigenvectors obtained from PCA of the SBgrid (left) and CLUMPY (right) model SED libraries.}
  \label{fig:pca_eigenvectors}
\end{figure*}

The first 16 eigenvectors that we have obtained for the two libraries are shown in Figure \ref{fig:pca_eigenvectors}.
This figure shows that low-order eigenvectors, which have larger variations in amplitude for different SEDs, are much smoother than the high-order eigenvectors,  which have smaller variations in amplitude for different SEDs.
The low-order eigenvectors determine the general shape of a SED while the high-order eigenvectors give some more details of the SED.
Then, any $\rm SED_i$, in the original library can be approximately reconstructed from a linear combination of these eigenvectors :
\begin{equation}
{\rm SED_i \approx \sum\limits_{j = 1}^{16} {{C_{i,j}}P{C_j}}}
\label{eq:pc2sed}
\end{equation}
Since the SEDs in the library have been normalized in advance, the corresponding inversions are needed after this.
An example of SED-Reconstruction is shown in Figure \ref{fig:testSED} of the next subsection to be compared with that obtained from ANN.
As \cite{AsensioRamos2009a}, we found that 16 PCs are enough to obtain an acceptable reconstruction for model SED libraries as complex as CLUMPY and SBgrid.
Now, each SED of the two libraries can be represented by a vector of 16 coefficients corresponding to 16  principal components (PCs), rather than a vector of 124 fluxes for CLUMPY library or a vector of 318 fluxes for SBgrid library.
It is clear that with the help of PCA the size of a model SED library can be greatly reduced.

\subsection{Implement of Artificial Neural Networks}
\label{ssect:ANN}
ANNs are mathematical constructs designed to simulate some intellectual behaviors of the human brain. 
For example, it can 'learn' relations between some inputs and outputs by training with many living examples.
After that, it can be used to predict the outputs from a new set of inputs.
Nowadays, ANNs have been used successfully in a wide range of problems in cosmology and astrophysics \citep{Lahav1996a,Bertin1996a,Andreon2000a,Firth2003a,Collister2004a,Vanzella2004a,Carballo2008a,Auld2008a}.
Here, we use ANNs to learn the relations between parameters and the resulting SEDs for libraries of model SEDs.
After training, an ANN can be used as a substitute to the model SED library which is used to train it, and even interpolate the library to obtain the SED for values of the parameters not present in the original grid.

There are different implements of ANNs which differ in the neurons (nodes) organization and information exchanging methods.
We have modified ANNz, a widely used tool for estimating photometric redshifts using ANN, to be a more convenient and general purpose ANN code without changing the technical implement of ANN. 
The type of ANN implemented in ANNz is so-called multi-layer perceptron (MLP) feed-forward network.
A  MLP  network consists of a number of layers of nodes with the first layer containing the inputs, the final layer containing the outputs, and one or more intervening (or ``hidden'')  layers.
In a feed-forward network, which is the most widely used due to its simplicity, information propagate sequentially from the input layer, through the hidden layers to the output neurons, without any feedback.
The network architecture of such an ANN can be denoted by $N_{\mathrm{in}}$:$N_1$:$N_2$: $\ldots$ :$N_{\mathrm{out}}$ where $N_{\mathrm{in}}$ is the number of input nodes, $N_i$ is the number of nodes in $i$th hidden layer, and $N_{\mathrm{out}}$ is the number of output nodes.

\begin{figure}
    \centering 
    \includegraphics[scale=0.2]{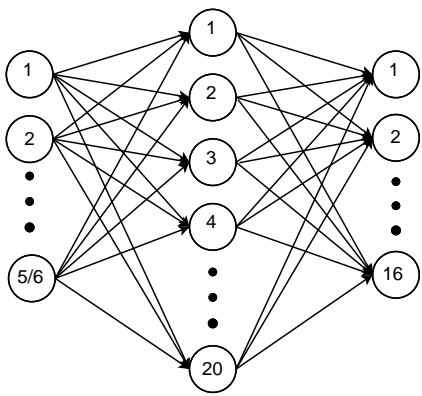} 
  \caption{The network architecture of ANNs for SBgrid (5 input parameters) and CLUMPY (6 input parameters) library.
  In both cases, a hidden layer with 20 nodes is used. The outputs of each ANN are the coefficients corresponding to the first 16 PCs.}
  \label{fig:ANN}
\end{figure}

In Figure \ref{fig:ANN}, we show the network architectures of ANNs used for SBgrid and CLUMPY libraries.
The inputs of an ANN are the parameters of the library used to train it.
So, the ANN for SBgrid library has 5 inputs while the ANN for CLUMPY library has 6 inputs.
The capability of an ANN is determined by the structure of its hidden layers.
In mathematics, the universal approximation theorem \citep{Cybenko1989a,Kurt1991a,Haykin1999a} states that a standard multilayer feed-forward network with only a single hidden layer and arbitrary continuous, bounded and nonconstant activation function can approximate any continuous function to arbitrary precision, provided only that sufficiently many hidden units are available.
More nodes in a single hidden layer or even more hidden layers can increase the degree of approximation, but with the expense of much more training time.
In practice, we found that a single hidden layer with 20 nodes is enough for the two libraries.
The outputs of ANNs are set to be the projections of a SED on the first 16 PCs (eigenvector).
So, the structure of ANNs for SBgrid and CLUMPY library can be denoted as 5:20:16 and 6:20:16, respectively.

An ANN ``learn'' the relationship between inputs and outputs from examples (pairs of inputs and corresponding outputs).
In our case, the examples are model SEDs of a library whose parameters and corresponding projections have already been known.
When a set of inputs is given, the ANN ``learn'' the relationship by adapting weights associated with connections between nodes so as to minimize the cost function, which represents the difference between the prediction of ANN and the expected outputs. 
An iterative quasi-Newtonian method is used in ANNz to perform this minimization.
Meanwhile, an activation  function, which is taken to be a sigmoid function in ANNz, is defined at each node  to simulate the behavior of biological neurons.
This defines the signal propagation rule of an ANN in the sense that a neuron is activated, which means it transmits the received signal further on,  when the total of received signals is greater than a certain threshold.

To avoid overfitting to the training set and optimize the generalization performance of the network, the SEDs in a library are separated into two sets: a training set and a validation set. 
Both of them are randomly selected from the library.
For the SBgrid library \footnote{Four SEDs in this library have been excluded, since they become discontinuous below about 1 $\micron$.}, the training set contains 6,495 ($90\%$) SEDs while the validation set contains 721 ($10\%$)  SEDs.
The CLUMPY library currently contains about 1307980 SEDs.
Although all of these SEDs can be used to train the ANN, we found that this is not necessary.
So, we have randomly selected about $10\%$ SEDs of the CLUMPY library, which contains 130,800 SEDs.
Then, 117,720 ($90\%$) of them are used as training set while the other 13,080 ($10\%$) are used as validation set.
On the other hand, the ANN usually converges to different local minima of the cost function, depending on the particular initialization.
So, for each library, a group of 4 networks (called a ``committee'') with the same structure but different initialization are trained independently, and the mean of the individual outputs of the 4 networks are used as a more accurate estimate for the outputs.

\begin{figure*}
  \plottwo{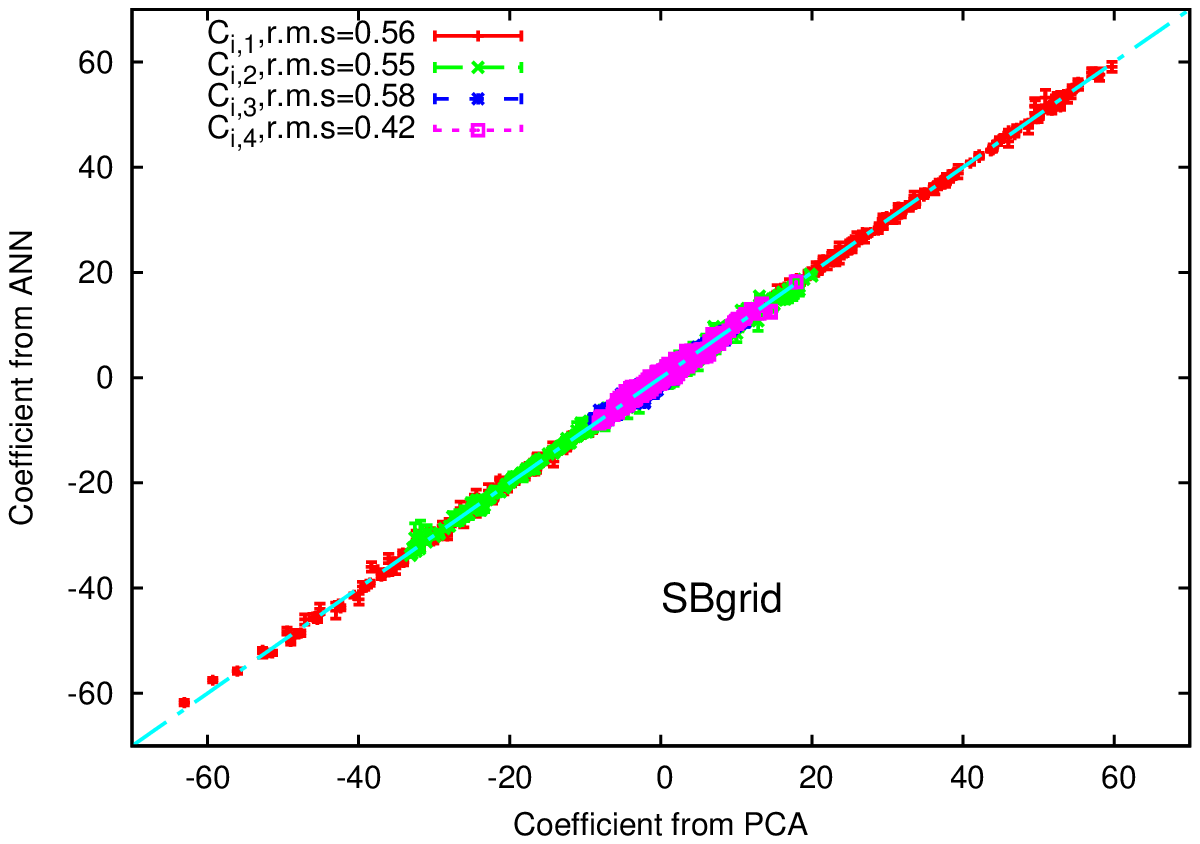}{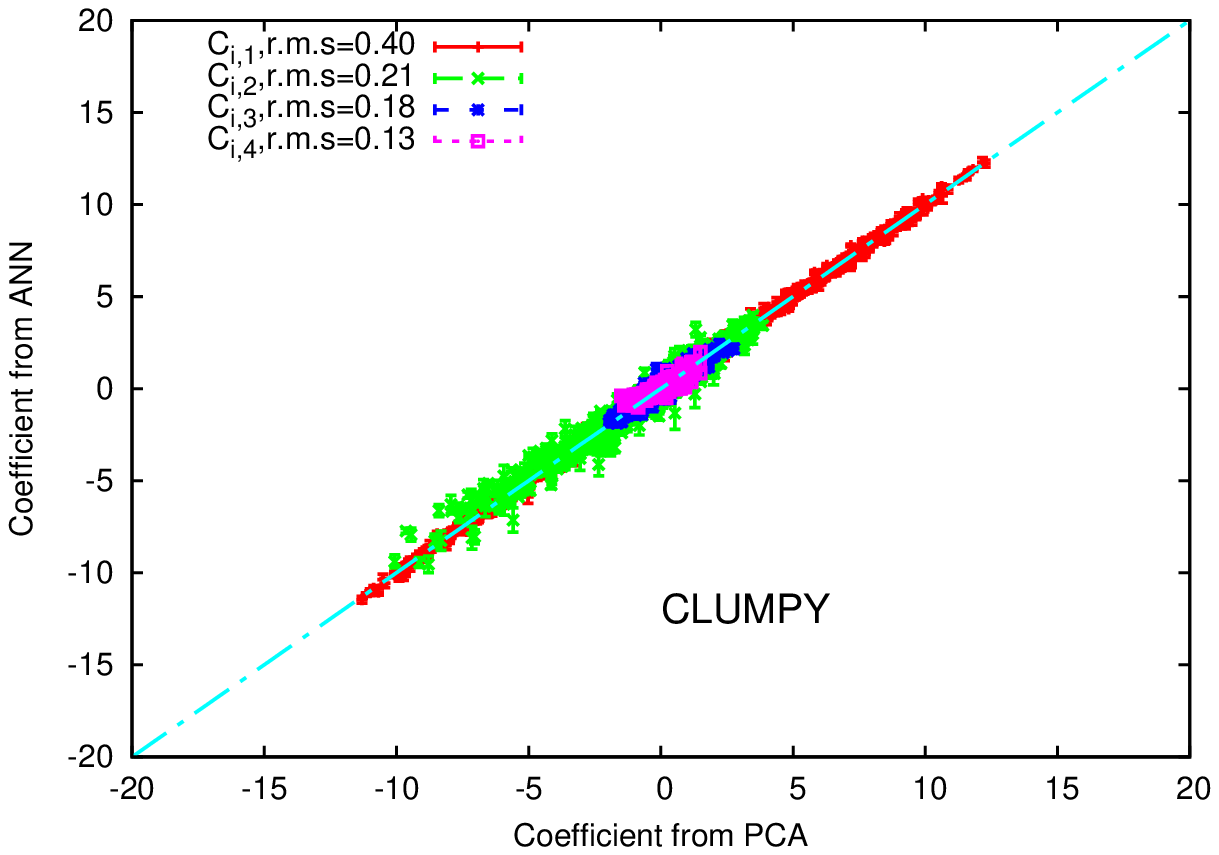}
  \caption{Left: The projections on the first 4 PCs for the 721 SEDs (validation set) randomly selected from the SBgrid library are predicted from ANN and then compared with that obtained directly from PCA.
  Right: As left, but for 13080 SEDs (validation set) randomly selected from the CLUMPY library.
  In both cases, the projections can be predicted from ANN with very small $\sigma_{\rm rms}$.
  So, the SEDs in the libraries can be reliably reconstructed from PCs by using these projections as coefficients for the linear combination of PCs.
  }
  \label{fig:testANN}
\end{figure*}
\begin{figure*}
  \plottwo{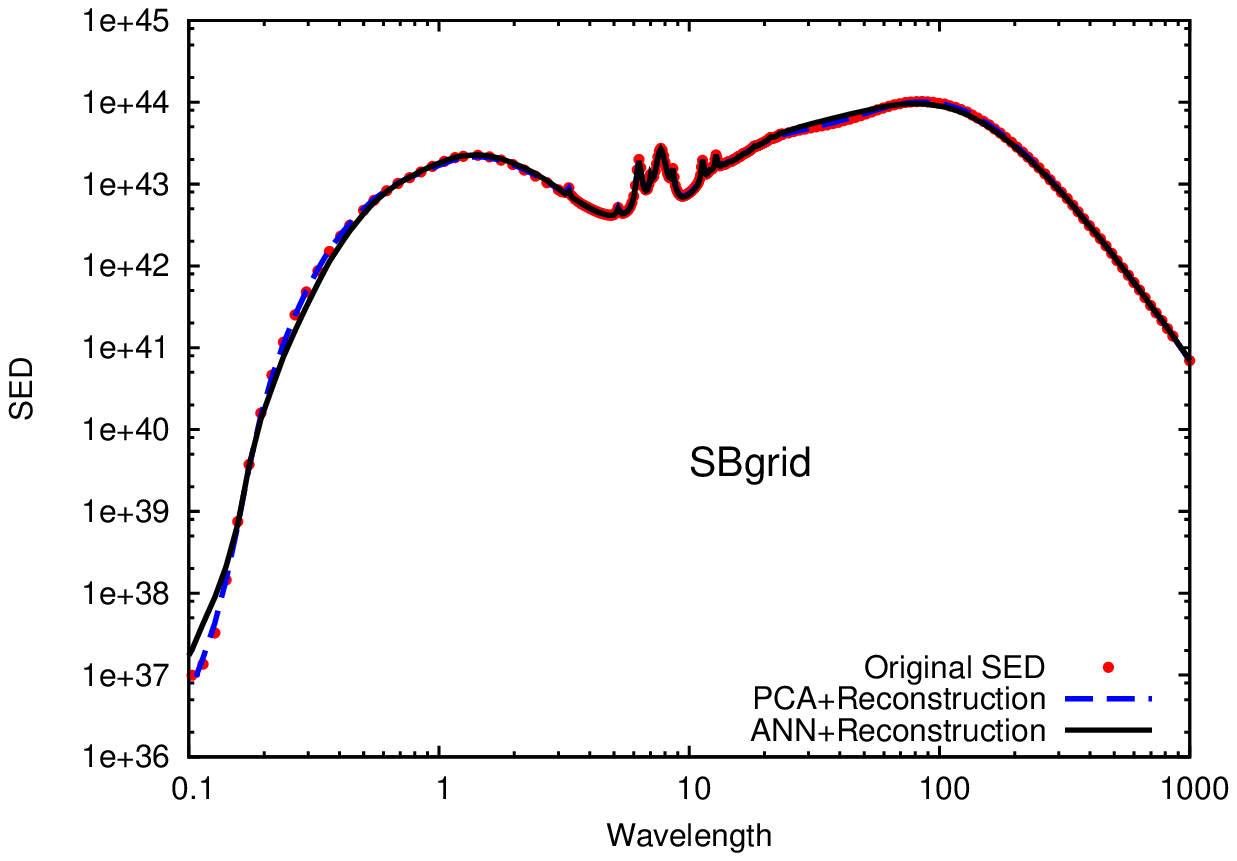}{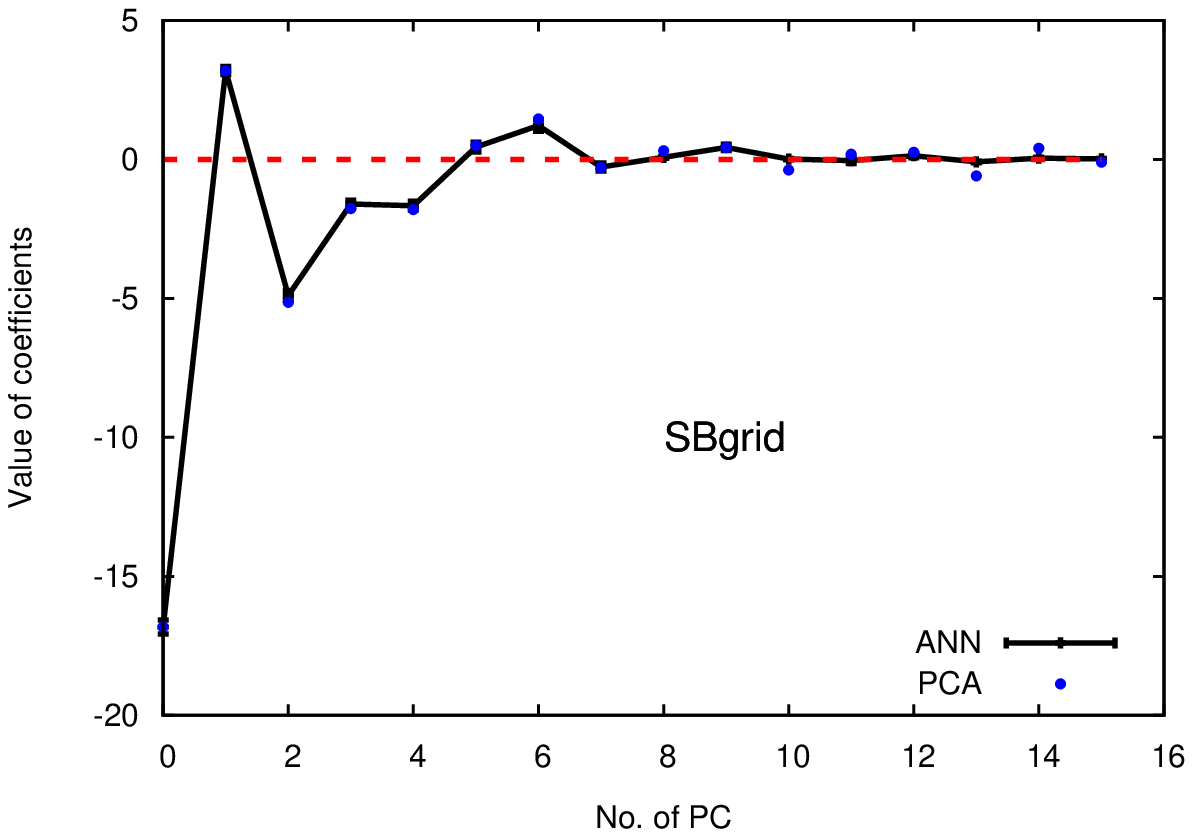}
  \plottwo{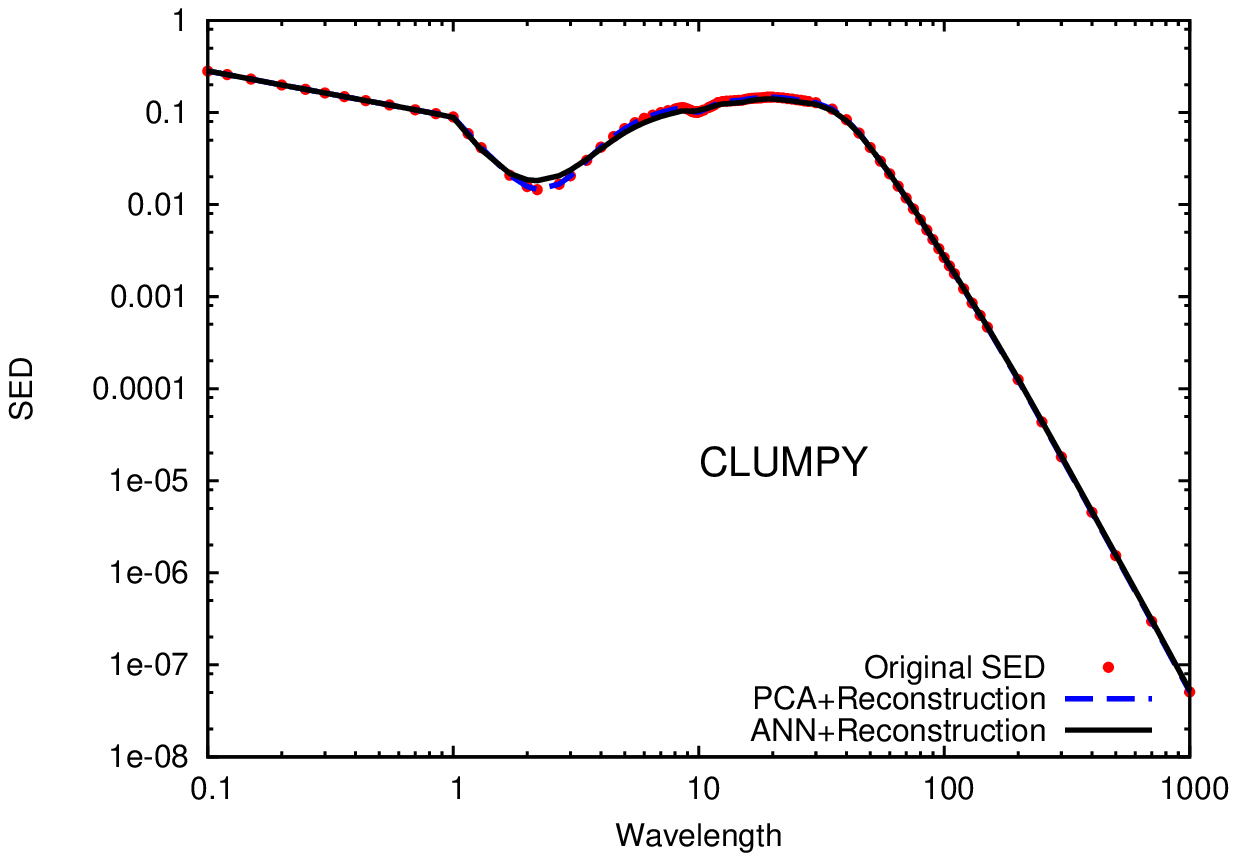}{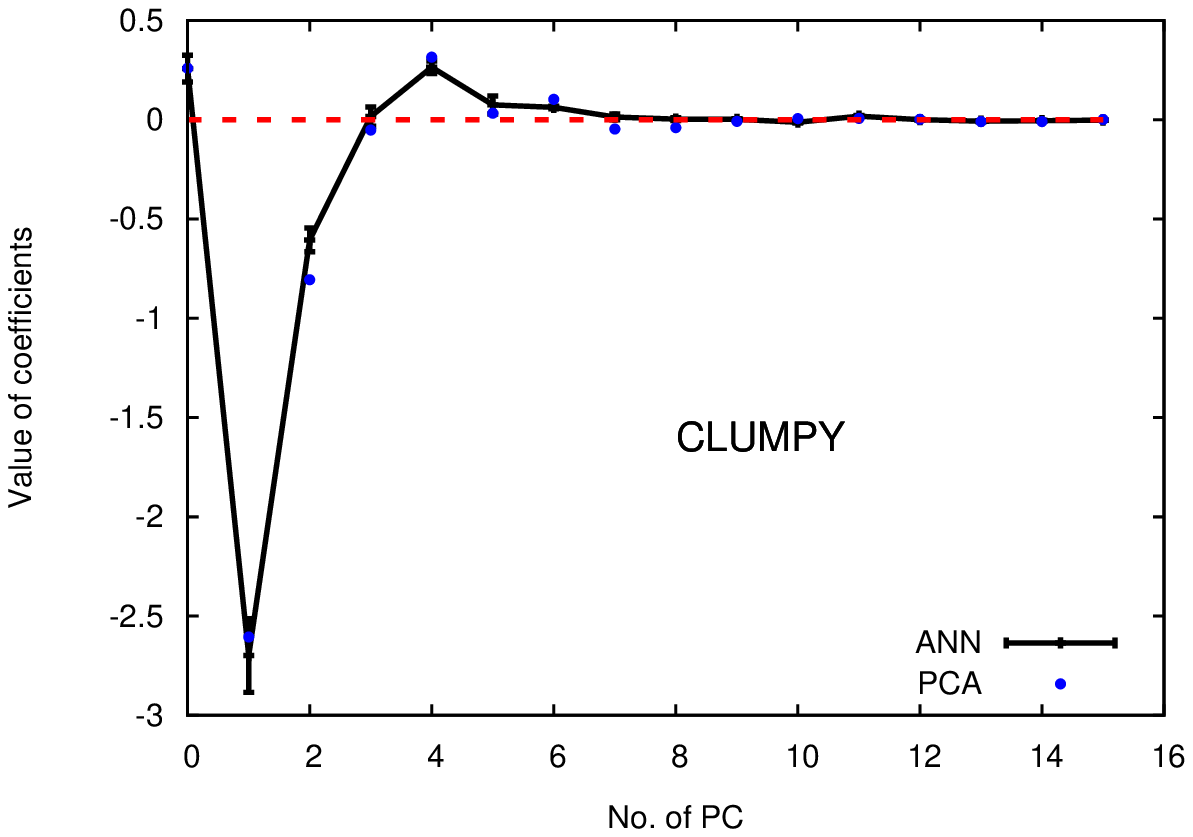}
  \caption{Left: Examples of model SED (red points) of SBgrid (top) and CLUMPY (bottom) library  compared with  that directly reconstructed by using the projections on the first 16 PCs (blue dash line) obtained by PCA of the library, and that reconstructed by using the projections predicted by ANN for the same set of parameters (black solid line). Right: The projections on the first 16 PCs of the model SED directly from PCA of the libraries (blue points) compared with that from ANN for the same set of parameters (black points with error bar and connected with black line).
  }
  \label{fig:testSED}
\end{figure*}

In Figure \ref{fig:testANN}, the projections on the first 4 PCs for the SEDs in the validation set from ANN are compared with that directly from PCA of the libraries.
As clearly shown, the projections can be reliably predicted from ANN.
For both libraries, the rms error $\sigma_{\rm rms}$ of the predicted projections are very small.
So, it is reasonable to expect that the SEDs in the libraries can be reliably reconstructed by using these projections as coefficients for the linear combination of PCs.
In Figure \ref{fig:testSED}, examples of SED-reconstruction by using projections as coefficients for the linear combination of PCs are shown.
It is clear that the SEDs in the two libraries can be reliably reconstructed by using the projections directly from PCA of the libraries or that predicted from ANN.

\section{BayeSED: A TOOL FOR BAYESIAN ANALYSIS OF SED}
\label{sect:BayeSED}
\subsection{Bayesian Inference}
\label{ssect:bayesian}
Bayesian methods have already been widely used in astrophysics and cosmology \citep[see, e.g.,][for a review]{Trotta2008a}. 
They have the advantage of higher efficiency and of a more consistent conceptual basis for dealing with problems  of induction in the presence of uncertainties than traditional statistical tools.
Bayesian methods are basically divided into two categories: parameter estimation and model comparison.
The basis of these methods is the so-called Bayes' Theorem which states that 
\begin{equation}
 \label{eq:Bayes_Theorem}
 P(\overrightarrow \theta|\overrightarrow d, M) = \frac{P(\overrightarrow d|\overrightarrow \theta, M) P(\overrightarrow \theta|M)}{P(\overrightarrow d|M)},
\end{equation}
where $\overrightarrow \theta$ represents a vector of parameters, $\overrightarrow d$ represents a vector of data sets, $M$ represents a model under consideration.

The left side of the Equation (\ref{eq:Bayes_Theorem}), $P(\overrightarrow \theta|\overrightarrow d, M)$ is called the {\em posterior probability} of parameter $\overrightarrow \theta$ given data $\overrightarrow d$ and model $M$.
It is proportional to the {\em sampling distribution} of the data $P(\overrightarrow d|\overrightarrow \theta,M)$ assuming the model is true, and the {\em prior probability} of the model, $P(\overrightarrow \theta|M)$ (``the prior''), which describes knowledges about the parameters acquired before seeing (or irrespective of) the data.
The {\em sampling distribution} describes how the degree of plausibility of the parameter $\overrightarrow \theta$ changes when new data $\overrightarrow d$ is acquired.
It is called {\em the likelihood function} when being considered as a function of the parameter $\overrightarrow \theta$, and often written as $L(\overrightarrow \theta) \equiv P(\overrightarrow d|\overrightarrow \theta,M)$.

The posterior  probability density function (PDF) for one parameter is obtained by marginalizing out (integrate out) other parameters from the full posterior distribution:
\begin{equation}
P(\theta_i|\overrightarrow d,M) = \int \mathrm{d}\theta_1 \cdots \mathrm{d}\theta_{i-1} \mathrm{d}\theta_{i+1} \cdots 
\mathrm{d}\theta_{N_\mathrm{par}} P(\overrightarrow \theta|\overrightarrow d,M).
\end{equation}
The normalization constant $P(\overrightarrow d|M)$, is called the {\em marginal likelihood} (or ``Bayesian evidence''), which is not important for parameter estimation but critical for model comparison, and given by
\begin{equation}
\label{eq:Bayes_evidence}
 P(\overrightarrow d|M) \equiv \sum_{\overrightarrow \theta}  P(\overrightarrow d|\overrightarrow \theta, M)
P(\overrightarrow \theta|M),
\end{equation}
where the sum runs over all the possible choices of the parameter $\overrightarrow \theta$.
For a continuous parameter space $\Omega_M$, this can be rewritten as:
\begin{equation}
\label{eq:evidence_def_mdl}
 P(\overrightarrow d | M) \equiv {\int_{\Omega_M} P(\overrightarrow d | \overrightarrow \theta, M)
 P(\overrightarrow \theta| M){\rm d}\overrightarrow \theta}.
\end{equation}

In the case of SED fitting, $\overrightarrow d$ represents the observed SED of a galaxy while $\overrightarrow \theta$ represents the parameters of a model SED library.
Commonly, $M$ represents a SED library as a whole.
However, multiple independent SED components (e.g., a starburst component and a AGN component) are needed in many cases.
Then, different combinations of independent SED components should be considered as different models.
All parameters of sub-models are combined together to be a new vector of parameters $\overrightarrow \theta$.
For libraries giving relative flux, a free scaling factor needs to be considered as an additional parameter in the new $\overrightarrow \theta$.

\subsection{Posterior Sampling Methods}
\label{ssect:sampling}
A key step in the Bayesian inference  problem outlined above is the evaluation of the posterior of Equation (\ref{eq:Bayes_Theorem}), where accurate analytical solutions are commonly not easy to obtain or just do not exist.
As a consequence, some efficient and robust sampling techniques have been developed.
A widely used technique is called Markov Chain Monte Carlo (MCMC).
An MCMC sampler, which is often based on the standard Metropolis-Hastings algorithm, provides such a way to explore the posterior distribution that the number density of samples is asymptotically converged to be proportional to the joint posterior PDF of all parameters.
So, it allows one to map out numerically the posterior distribution even in the case where the parameter space has hundreds of dimensions and the posterior is multimodal and with a complicated structure.

However, such methods can be very computationally intensive when the posterior distribution is multimodal or with large degeneracies between parameters, particularly in high dimensions.
On the other hand, the calculation of Bayesian evidence, which is the key ingredient for Bayesian model comparison, is extremely computationally intensive by using MCMC techniques. 
Another Monte Carlo method called Nested sampling \citep{Skilling2004a,Mukherjee2006a,Shaw2007a} provides a more efficient method for the calculation of Bayesian evidence, but also produces posterior inferences as a by-product. 
Here, we adopt a newly developed, highly efficient and freely available Bayesian inference tool, called {\sc MultiNest} \citep{Feroz2008a,Feroz2009a}.
It is as efficient as standard MCMC methods for Bayesian parameter estimation, more efficient for very accurate evaluation of Bayesian evidences for model comparison, and fully parallelized.

\subsection{Building-up of BayeSED}
\label{ssect:BayeSED}
\begin{figure}
    \centering 
    \includegraphics[scale=0.4]{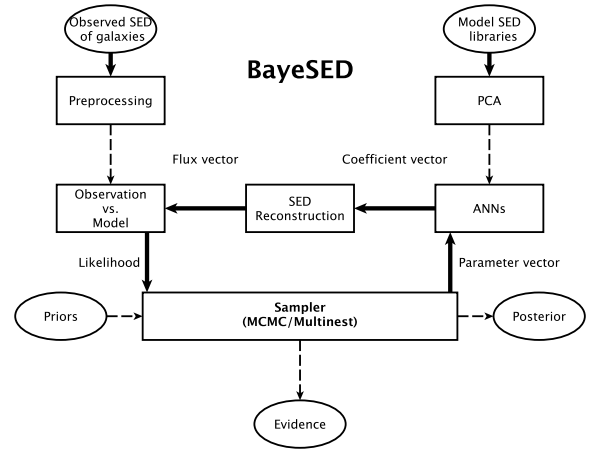} 
  \caption{A simple flowchart for the Bayesian analysis of SED boosted by PCA and ANN.
  }
  \label{fig:flowchart}
\end{figure}
The general structure of our Bayesian inference tool for the analysis of SED, BayeSED, is shown in Figure \ref{fig:flowchart}.
The core of BayeSED is the sampling of posterior probability with a MCMC or MultiNest sampler.
This is shown as a loop in the figure.
During the sampling, the sampler provides proposal parameter vectors for the ANN, and the ANN predicts the coefficients for the reconstruction of model SEDs using the proposed parameter set.
After a training with some model SEDs in a library, an ANN can help to generate the model SED of any parameter vector within the parameter space covered by the library used to trained it.
Here, it is allowed to simultaneously use multiple ANNs, which are trained with different model SED libraries.
The comparison of model with observations gives a ${\chi(\overrightarrow \theta)}^2$.
Then, the likelihood function is given by $L(SED_{\rm obs}|\overrightarrow \theta,M)\equiv{e^{ - {\chi(\overrightarrow \theta)}^2/2}}$.

The priors represent our knowledges about the parameters of the model that are independent of current observations.
If we have no prior knowledge about the model parameters, the prior distributions are commonly assumed to be uniform  between two physically chosen bounds.
When the sampling is converged, the posterior PDF for all parameters and the Bayesian evidence for the model are obtained.
Then, a posterior distribution differing from a uniform distribution would imply that new information about the corresponding parameter are obtained from observations.
On the other hand, the ratio of evidences for two models, the so-called ``Bayes factor'', tells us how their relative plausibility  should be changed as suggested by the new observations.

\section{Application to a sample of hyperluminous infrared galaxies}
\label{sect:application}
\subsection{The HLIRG sample and data}
\label{ssect:sample}
The sample studied here is the one selected by \cite{Ruiz2007a} from the \cite{Rowan-Robinson2000a} sample of 45 HLIRGs.
The sample is limited to sources with available X-ray data and with redshift less than $\sim2$ to avoid strong biasing towards high redshift quasars.
Consequently, the final sample contains thirteen objects.
\cite{Ruiz2010a} have built multi-wavelength (from radio to X-rays) SEDs for these HLIRGs.
They fitted standard empirical AGN and starburst templates to these SEDs and classified the HLIRGs into two groups, named class A and class B, according to their different SED shapes.
These authors also suggested that their simple template-fitting approach should be complemented with other theoretical models of starburst and AGN emission.

Here, we present a re-analysis of the SEDs of these HLIRGs by using different RT models of starburst and AGN emission, and put it on a solid statistical basis.
The redshifts and observed SEDs of these galaxies are taken from the Table 1 and B of \cite{Ruiz2010a}, respectively.
The SEDs have been converted to monochromatic flux density, corrected for the Galactic reddening and blue-shifted to rest-frame.
Before comparing with model SEDs, we convert the monochromatic flux density to monochromatic luminosity by using the luminosity distance $d_L(z)$.

\subsection{Bayesian analysis of SEDs}
\label{ssect:analysis}
\subsubsection{Models and priors}
\label{sssect:models}
Three different models are employed to do Bayesian analysis of the SEDs of these HLIRGs.
The first is the pure starburst model of \cite{Siebenmorgen2007a} as presented in the SBgrid library (noted as `SB' hereafter).
The priors for the 5 parameters of this model are assumed to be uniform  distributions truncated to  the following intervals: $R =[0.35,15]$ kpc, $f_{\rm OB} = [0.4,1]$, $\rm log(L_{\rm tot}/L_{\odot}) = [10,14.7]$ , $A_{\rm V} = [2.2,144]$, $\rm log(n_{\rm hs}/cm^{-3}) = [2,4]$.
The SEDs in the SBgrid library are in unit of absolute flux at a distance of $\rm 50Mpc$.
The absolute flux values have been multiplied by $\rm 4\pi*50Mpc*50Mpc$ in advance to convert to absolute luminosity values.

The second is the pure AGN model of \cite{Nenkova2008a} as presented in the CLUMPY library (noted as `AGN' hereafter).
The priors for the 6 parameters  of this model  are also assumed to be uniform distributions truncated to  the following intervals: $\sigma = [15,75]$, $Y=[5,200]$, $N = [1,24]$, $q=[0,4.5]$, $\tau_{\rm V} = [5,200]$, $i = [0,90]$.
Since the SEDs in the CLUMPY library have been normalized, an additional scaling factor needs to be considered as a new parameter.
The prior for this parameter is assumed to be uniform in the log space: $\rm log(scale_{AGN}/erg\;s^{-1})$ = [44,50].

Finally, the linear combination of the pure starburst and pure AGN models is considered as an additional new model (noted as `SB+AGN' hereafter).
The assumed priors are the same ones as above.
As discussed in \S \ref{ssect:ANN}, the model SEDs in the two libraries have been used to train two groups of ANNs, respectively. 
The trained ANNs are used as substitutions of the original models, and the models can be evaluated continuously in the whole parameter space.
Since both of the starburst and AGN model used here are not extended to the X-ray range, in this paper we mainly focus on the IR range (i.e. $\rm 1-1000~\mu m$) of the SEDs.
The X-ray data for galaxies in the HLIRGs sample have also been provided by \cite{Ruiz2010a}.
However, it is very hard to construct a self consistent SED model that is able to reproduce the whole SED covering such a wide range of wavelengths.

\subsubsection{Model comparison}
\label{sssect:evidences}
The  Bayesian  evidence represents a practical implementation of the Occam’s razor principle.
So, a complex model with more parameters has a lower Bayesian evidence unless it provides a significantly better fitting to the observations.
As mentioned above, in this paper we consider three different models: `SB', `AGN', and `SB+AGN'.
They have 5, 7 and 12 parameters, respectively.
\begin{table}
\caption{The Bayesian evidences of the `SB', `AGN', and `SB+AGN' models for galaxies in the class A and B of \cite{Ruiz2007a} HLIRGs sample.}
\centering
\scalebox{0.9}
{
\begin{tabular}{llll}
\hline\hline\noalign{\smallskip}
Source&$\mathrm{ln(ev_{SB})}$&$\mathrm{ln(ev_{AGN})}$&$\mathrm{ln(ev_{SB+AGN})}$\\
\hline\noalign{\smallskip}
Class A HLIRG\\
PG1206+459&$\mathrm{-12.65^{+0.09}_{-0.09}}$&$\mathrm{-15.08^{+0.08}_{-0.08}}$&$\mathrm{-9.39^{+0.08}_{-0.08}}$\\
PG1247+267&$\mathrm{-11.45^{+0.09}_{-0.09}}$&$\mathrm{-8.02^{+0.08}_{-0.08}}$&$\mathrm{-7.71^{+0.08}_{-0.08}}$\\
IRASF12509+3122&$\mathrm{-9.78^{+0.09}_{-0.09}}$&$\mathrm{-7.81^{+0.07}_{-0.07}}$&$\mathrm{-7.38^{+0.07}_{-0.07}}$\\
IRAS14026+4341&$\mathrm{-54.16^{+0.13}_{-0.13}}$&$\mathrm{-21.64^{+0.12}_{-0.12}}$&$\mathrm{-19.27^{+0.12}_{-0.12}}$\\
IRASF14218+3845&$\mathrm{-5.91^{+0.06}_{-0.06}}$&$\mathrm{-6.36^{+0.06}_{-0.06}}$&$\mathrm{-5.59^{+0.06}_{-0.06}}$\\
IRAS16347+7037&$\mathrm{-35.98^{+0.12}_{-0.12}}$&$\mathrm{-23.48^{+0.12}_{-0.12}}$&$\mathrm{-18.10^{+0.12}_{-0.12}}$\\
IRAS18216+6418&$\mathrm{-172.25^{+0.14}_{-0.14}}$&$\mathrm{-72.53^{+0.14}_{-0.14}}$&$\mathrm{-26.36^{+0.14}_{-0.14}}$\\
Class B HLIRG\\
IRASF00235+1024&$\mathrm{-11.15^{+0.09}_{-0.09}}$&$\mathrm{-38.92^{+0.09}_{-0.09}}$&$\mathrm{-11.09^{+0.09}_{-0.09}}$\\
IRAS07380-2342&$\mathrm{-159.50^{+0.12}_{-0.12}}$&$\mathrm{-179.24^{+0.15}_{-0.15}}$&$\mathrm{-76.36^{+0.17}_{-0.17}}$\\
IRAS00182-7112&$\mathrm{-22.09^{+0.10}_{-0.10}}$&$\mathrm{-57.88^{+0.14}_{-0.14}}$&$\mathrm{-19.10^{+0.12}_{-0.12}}$\\
IRAS09104+4109&$\mathrm{-31.31^{+0.11}_{-0.11}}$&$\mathrm{-71.56^{+0.15}_{-0.15}}$&$\mathrm{-29.62^{+0.13}_{-0.13}}$\\
IRAS12514+1027&$\mathrm{-63.33^{+0.10}_{-0.10}}$&$\mathrm{-62.12^{+0.13}_{-0.13}}$&$\mathrm{-30.01^{+0.14}_{-0.14}}$\\
IRASF15307+3252&$\mathrm{-12.64^{+0.10}_{-0.10}}$&$\mathrm{-51.24^{+0.14}_{-0.14}}$&$\mathrm{-11.97^{+0.10}_{-0.10}}$\\
\noalign{\smallskip}\hline\noalign{\smallskip}
\end{tabular}

}
\label{tab:ev}
\end{table}
In Table~\ref{tab:ev}, we present the Bayesian evidences of the three models for HLIRGs in the class A and class B as defined by \cite{Ruiz2010a}.
Since the Bayesian evidences for different models spread in a very wide range, we use $\rm ln(ev_{model})$ instead of the evidence itself.

As shown in Table~\ref{tab:ev}, the `SB+AGN' model has the highest Bayesian evidence for all galaxies in the HLIRGs sample, although it has the largest number of parameters.
So, `SB+AGN' model provides a much better fitting to all of these HLIRGs, which means starburst and AGN activities are probably ongoing together in these galaxies.
On the other hand, for most class A HLIRGs the pure `AGN' model has a higher Bayesian evidence than a pure `SB', while for most class B HLIRGs the pure `SB' model has a higher Bayesian evidence than a pure `AGN' model.
These results imply that class A HLIRGs are dominated by AGN while class B HLIRGs are dominated by starburst, although both of starburst and AGN are present in all cases.

\subsubsection{Parameter estimation}
\label{sssect:parameters}

\begin{figure*}
\centering 
\plotone{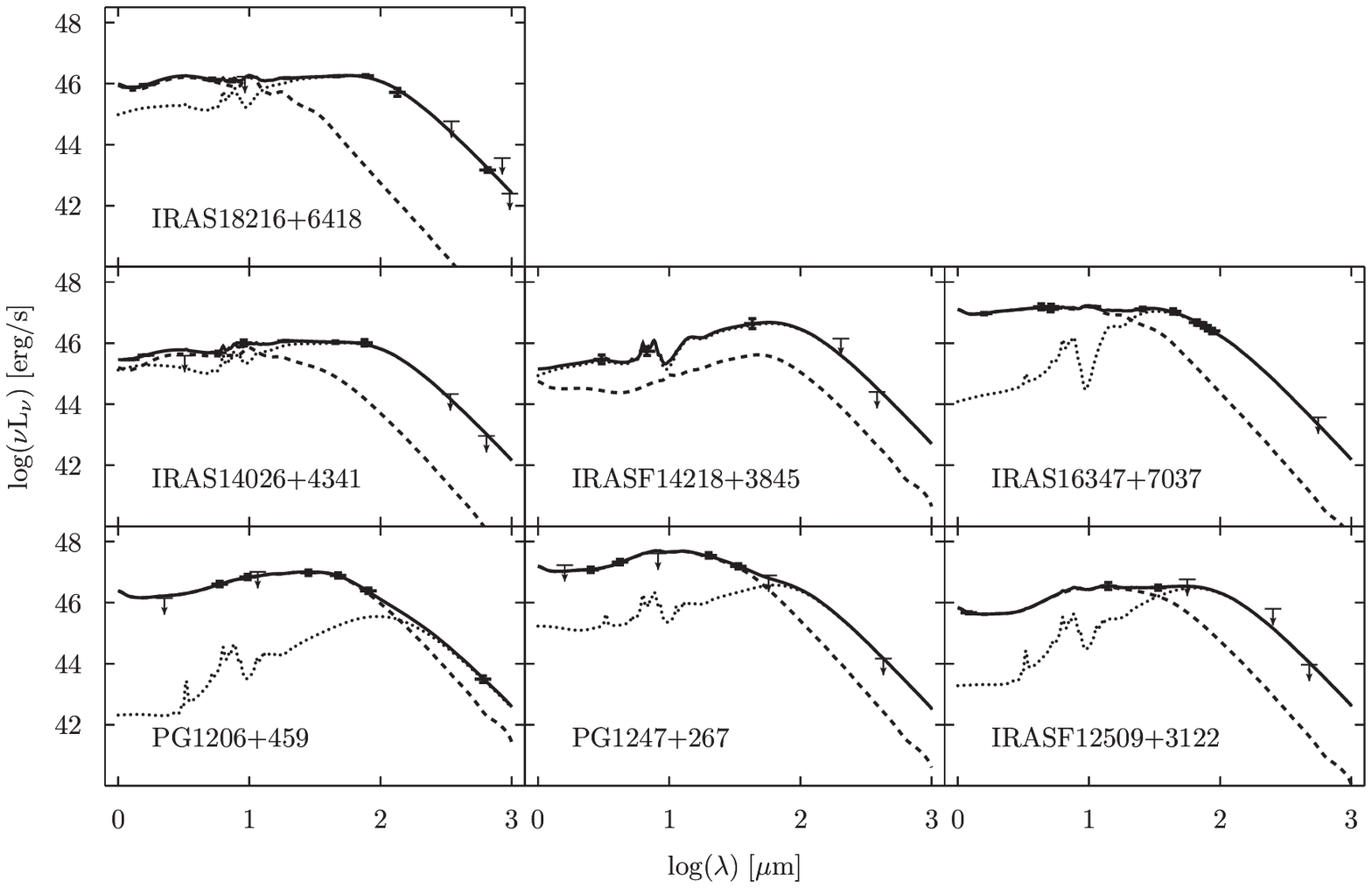}
\caption{Best fit (or MAP) model SEDs for class A HLIRG obtained from sampling the `SB+AGN' model, which has the highest Bayesian evidence among the three models considered.
The dotted, dashed, and solid lines  represent the starburst component, AGN component and total, respectively.}
\label{fig:MAP_modeA}
\end{figure*}

\begin{figure*}
\centering 
\plotone{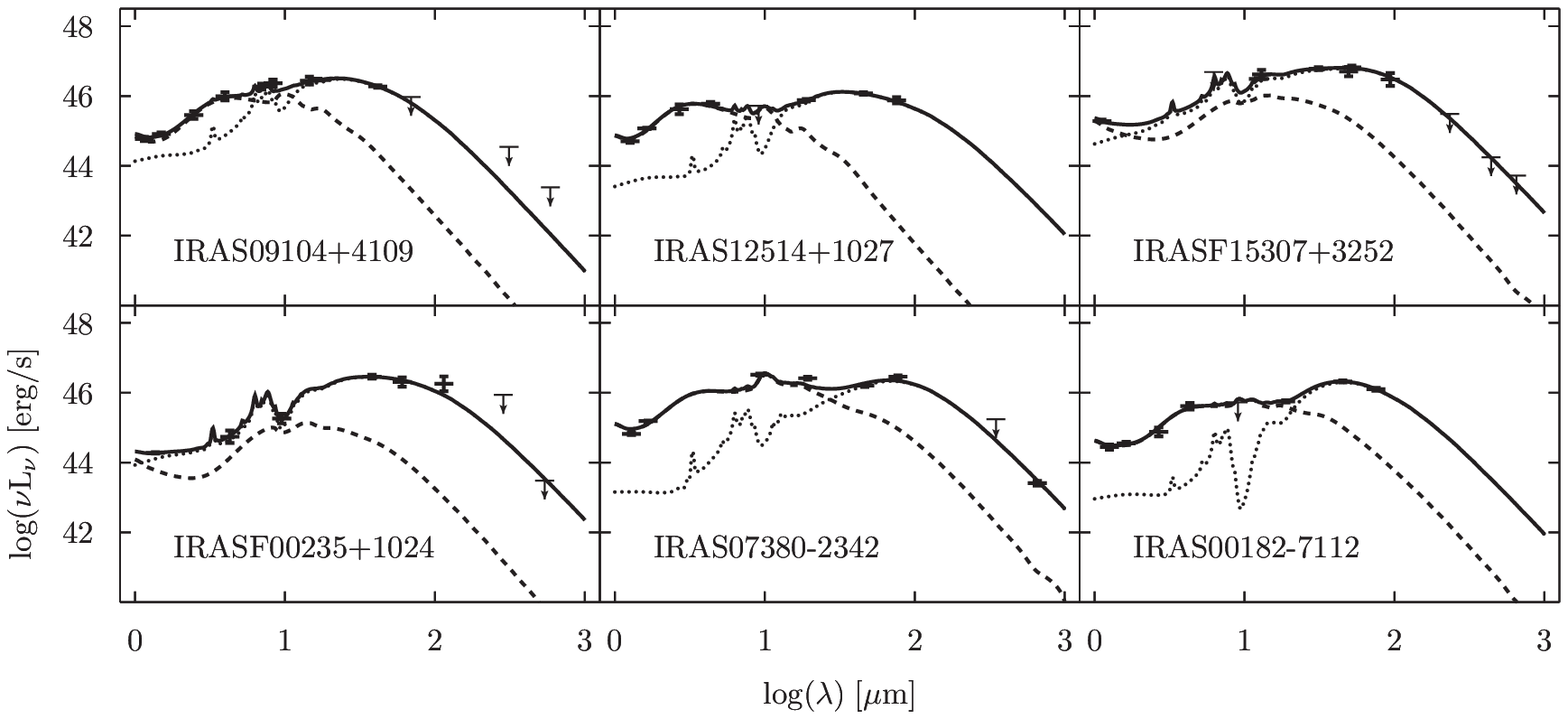}
\caption{Similar to Figure \ref{fig:MAP_modeA}, but for class B HLIRG.}
\label{fig:MAP_modeB}
\end{figure*}

In Figure \ref{fig:MAP_modeA} and \ref{fig:MAP_modeB}, we show the best fit i.e. the maximum a posteriori (MAP) model SEDs that are found during sampling parameter space of the `SB+AGN' model, which has the highest Bayesian evidence among the three models considered.
Commonly, the values of parameters corresponding to these best models are taken as the best estimation of parameters.
However, the Bayesian analysis method  has the advantage of providing the detailed posterior distributions for all parameters, which represent our full knowledge about these parameters given the priors and new observations.
The best expectations and uncertainties about all parameters can be deduced from these possibility distributions.

\begin{figure*}
  \centering 
  \plotone{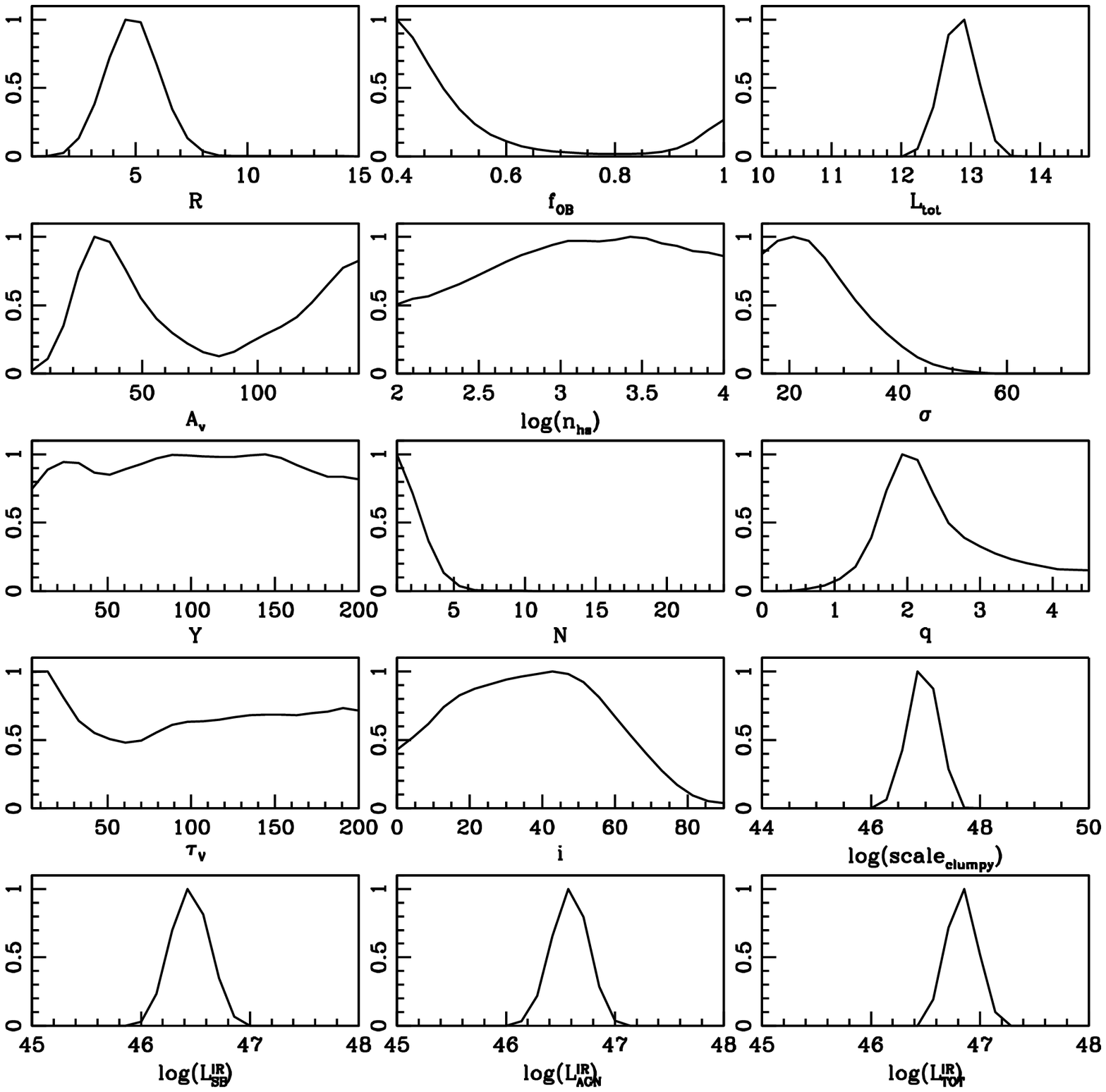} 
  \caption{
  One-dimensional marginal posterior probability density functions of the basic and derived parameters of `SB+AGN' model for IRAS18216+6418.
  It is clear that some parameters are poorly constrained.
  However, the derived IR luminosities are nicely constrained.
  }
  \label{fig:iras18216SBclumpy}
\end{figure*}

In Figure \ref{fig:iras18216SBclumpy}, we show the posterior PDFs of all parameters of the `SB+AGN' model for IRAS18216+6418.
Given the very limited observations and the large number of parameters, it is clear that not all of the parameters can be well constrained.
From the detailed posterior PDFs of parameters, it is much easy to find out if a parameter is well constrained or not.
For example, the basic parameters $R$, $L_{\rm tot}$ of starburst component and $\sigma$, $q$ of AGN component are well constrained.
The derived parameters $\rm log(L_{SB}^{IR})$, $\rm log(L_{AGN}^{IR})$ and $\rm log(L_{TOT}^{IR})$ are nicely constrained.

Apparently, for a large number of galaxies in a sample, it is not possible to plot such kind of PDFs for all of them.
It would be more convenient to use a summary statistics to give a good estimate for a parameter and its spread.
Here, we use the median and percentile statistics.
The median are found by firstly sorting all values in ascending order, then taking the element in the middle so that half of all points are below the median and the other half above it.
The lower and upper quartiles are the values below which 25 and 75 percent of points fall, respectively.
This statistics is much better than the mostly used mean and standard deviation statistics when the distribution of PDF is asymmetrically skewed or multimodal.

\subsubsection{Relations between starburst and AGN parameters}
\label{sssect:relations}
\begin{table*}
  \centering
  \caption{The estimated starburst parameters and corresponding uncertainties  for class A and B HLIRGs by employing the `SB+AGN' model.
The median and percentile statistics are used to the obtain the best estimation of a parameter and its upper and lower limits.
}
  \begin{tabular}{lllllll}
\hline\hline\noalign{\smallskip}
Source&$\mathrm{R}$&$\mathrm{f_{OB}}$&$\mathrm{L_{tot}}$&$\mathrm{A_v}$&$\mathrm{log(n_{hs})}$&$\mathrm{log(L_{SB}^{IR})}$\\
&$\mathrm{(kpc)}$&&$\mathrm{(L_{\odot})}$&$\mathrm{(mag)}$&$\mathrm{(cm^{-3})}$&$\mathrm{(erg/s)}$\\
\hline\noalign{\smallskip}
Class A HLIRG\\
PG1206+459&$6.99^{+3.07}_{-2.14}$&$0.72^{+0.13}_{-0.14}$&$12.19^{+0.73}_{-0.99}$&$67.83^{+34.71}_{-31.51}$&$3.00^{+0.48}_{-0.48}$&$46.25^{+0.46}_{-0.59}$\\
PG1247+267&$7.66^{+3.38}_{-3.25}$&$0.69^{+0.15}_{-0.14}$&$11.98^{+0.95}_{-0.89}$&$69.44^{+34.67}_{-34.21}$&$3.01^{+0.47}_{-0.49}$&$46.13^{+0.69}_{-0.60}$\\
IRASF12509+3122&$7.50^{+2.97}_{-2.75}$&$0.67^{+0.15}_{-0.13}$&$12.59^{+0.49}_{-0.98}$&$47.93^{+40.44}_{-27.90}$&$3.01^{+0.47}_{-0.47}$&$46.47^{+0.31}_{-0.60}$\\
IRAS14026+4341&$5.28^{+2.39}_{-1.27}$&$0.62^{+0.18}_{-0.13}$&$12.16^{+0.55}_{-0.94}$&$33.03^{+46.58}_{-14.07}$&$3.01^{+0.48}_{-0.46}$&$46.18^{+0.17}_{-0.52}$\\
IRASF14218+3845&$6.47^{+2.89}_{-2.25}$&$0.63^{+0.17}_{-0.13}$&$12.94^{+0.30}_{-1.13}$&$63.27^{+36.66}_{-27.43}$&$2.99^{+0.49}_{-0.47}$&$46.69^{+0.17}_{-0.59}$\\
IRAS16347+7037&$3.58^{+2.89}_{-1.19}$&$0.71^{+0.13}_{-0.14}$&$13.42^{+0.19}_{-1.02}$&$69.83^{+22.19}_{-23.45}$&$2.91^{+0.49}_{-0.43}$&$47.08^{+0.11}_{-0.57}$\\
IRAS18216+6418&$5.08^{+1.09}_{-0.89}$&$0.54^{+0.22}_{-0.10}$&$12.78^{+0.10}_{-0.12}$&$60.16^{+48.19}_{-26.65}$&$3.07^{+0.45}_{-0.48}$&$46.41^{+0.11}_{-0.08}$\\
Class B HLIRG\\
IRASF00235+1024&$5.02^{+1.88}_{-1.33}$&$0.86^{+0.07}_{-0.12}$&$12.98^{+0.06}_{-0.25}$&$59.78^{+35.62}_{-20.74}$&$2.96^{+0.50}_{-0.47}$&$46.64^{+0.06}_{-0.14}$\\
IRAS07380-2342&$8.32^{+1.13}_{-1.36}$&$0.88^{+0.09}_{-0.13}$&$12.82^{+0.06}_{-0.06}$&$47.66^{+31.68}_{-15.95}$&$3.25^{+0.46}_{-0.55}$&$46.53^{+0.05}_{-0.05}$\\
IRAS00182-7112&$2.84^{+1.54}_{-0.84}$&$0.74^{+0.12}_{-0.13}$&$12.86^{+0.06}_{-0.07}$&$84.62^{+14.86}_{-14.57}$&$2.95^{+0.48}_{-0.46}$&$46.46^{+0.03}_{-0.03}$\\
IRAS09104+4109&$1.21^{+2.88}_{-0.25}$&$0.91^{+0.03}_{-0.06}$&$13.11^{+0.05}_{-0.07}$&$21.13^{+12.85}_{-7.49}$&$3.42^{+0.37}_{-0.55}$&$46.70^{+0.06}_{-0.04}$\\
IRAS12514+1027&$3.89^{+1.20}_{-1.07}$&$0.83^{+0.09}_{-0.15}$&$12.64^{+0.05}_{-0.07}$&$49.66^{+17.00}_{-11.59}$&$3.07^{+0.47}_{-0.50}$&$46.30^{+0.04}_{-0.07}$\\
IRASF15307+3252&$7.02^{+1.33}_{-1.44}$&$0.72^{+0.10}_{-0.09}$&$13.47^{+0.05}_{-0.09}$&$27.28^{+21.99}_{-8.87}$&$3.11^{+0.45}_{-0.50}$&$47.12^{+0.04}_{-0.09}$\\
\noalign{\smallskip}\hline\noalign{\smallskip}
\end{tabular}

  \label{tab:SB}
\end{table*}

\begin{table*}
  \centering
  \caption{Similar to Table~\ref{tab:SB}, but for AGN parameters.}
  \begin{tabular}{llllllll}
\hline\hline\noalign{\smallskip}
Source&$\mathrm{\sigma}$&$\mathrm{Y}$&$\mathrm{N}$&$\mathrm{q}$&$\mathrm{\tau_V}$&$\mathrm{i}$&$\mathrm{log(L_{AGN}^{IR})}$\\
&&&&&&&$\mathrm{(erg/s)}$\\
\hline\noalign{\smallskip}
Class A HLIRG\\
PG1206+459&$40.05^{+12.56}_{-11.22}$&$97.48^{+44.47}_{-42.53}$&$11.11^{+5.36}_{-4.63}$&$1.85^{+0.76}_{-0.66}$&$111.98^{+37.44}_{-36.20}$&$31.94^{+22.61}_{-16.85}$&$47.31^{+0.06}_{-0.21}$\\
PG1247+267&$34.70^{+14.38}_{-10.24}$&$93.72^{+45.77}_{-43.87}$&$9.72^{+5.54}_{-4.18}$&$3.13^{+0.68}_{-0.90}$&$81.23^{+48.02}_{-27.97}$&$36.34^{+20.79}_{-17.24}$&$47.85^{+0.10}_{-0.41}$\\
IRASF12509+3122&$41.73^{+13.82}_{-11.81}$&$99.25^{+47.14}_{-45.48}$&$12.57^{+5.26}_{-5.05}$&$2.40^{+0.87}_{-0.86}$&$97.81^{+46.83}_{-44.13}$&$41.77^{+20.61}_{-19.82}$&$46.74^{+0.11}_{-0.24}$\\
IRAS14026+4341&$26.93^{+6.37}_{-5.43}$&$149.83^{+31.80}_{-57.93}$&$8.17^{+4.95}_{-4.57}$&$2.23^{+0.68}_{-0.41}$&$151.33^{+22.62}_{-28.80}$&$31.04^{+7.95}_{-8.56}$&$46.45^{+0.08}_{-0.22}$\\
IRASF14218+3845&$39.59^{+14.20}_{-12.35}$&$109.19^{+43.97}_{-45.32}$&$11.79^{+5.87}_{-5.67}$&$1.70^{+1.13}_{-0.81}$&$113.35^{+41.27}_{-45.08}$&$42.45^{+21.09}_{-19.50}$&$46.21^{+0.39}_{-0.38}$\\
IRAS16347+7037&$23.22^{+7.87}_{-4.85}$&$133.65^{+36.68}_{-49.48}$&$3.28^{+4.31}_{-1.41}$&$2.90^{+0.71}_{-0.80}$&$80.42^{+57.14}_{-33.09}$&$45.90^{+13.40}_{-15.82}$&$47.56^{+0.04}_{-0.03}$\\
IRAS18216+6418&$22.75^{+6.87}_{-4.27}$&$128.43^{+41.11}_{-55.38}$&$2.27^{+2.07}_{-0.75}$&$2.26^{+0.76}_{-0.40}$&$127.57^{+38.67}_{-48.49}$&$41.81^{+14.00}_{-16.55}$&$46.63^{+0.05}_{-0.09}$\\
Class B HLIRG\\
IRASF00235+1024&$47.49^{+13.68}_{-14.45}$&$99.18^{+46.82}_{-45.71}$&$13.55^{+5.06}_{-5.49}$&$2.72^{+0.94}_{-1.20}$&$100.74^{+47.47}_{-45.73}$&$42.04^{+22.69}_{-20.55}$&$45.36^{+0.17}_{-0.17}$\\
IRAS07380-2342&$34.96^{+7.86}_{-3.97}$&$178.88^{+16.22}_{-93.19}$&$18.85^{+2.71}_{-9.16}$&$0.65^{+2.91}_{-0.59}$&$58.29^{+126.95}_{-52.79}$&$15.56^{+8.71}_{-5.52}$&$46.39^{+0.20}_{-0.24}$\\
IRAS00182-7112&$45.54^{+11.20}_{-11.03}$&$109.72^{+44.76}_{-48.92}$&$12.53^{+4.94}_{-4.97}$&$2.77^{+0.82}_{-1.04}$&$27.35^{+81.14}_{-15.69}$&$30.07^{+19.92}_{-14.24}$&$45.67^{+0.13}_{-0.22}$\\
IRAS09104+4109&$43.17^{+11.33}_{-8.95}$&$96.12^{+54.40}_{-57.72}$&$14.38^{+4.72}_{-5.76}$&$2.75^{+0.82}_{-1.25}$&$39.52^{+70.69}_{-31.15}$&$28.05^{+19.02}_{-12.07}$&$45.95^{+0.19}_{-0.37}$\\
IRAS12514+1027&$42.50^{+7.13}_{-7.84}$&$124.10^{+50.15}_{-75.66}$&$6.08^{+4.86}_{-1.84}$&$2.90^{+0.72}_{-1.19}$&$29.47^{+123.27}_{-23.40}$&$21.00^{+9.78}_{-8.41}$&$45.96^{+0.09}_{-0.07}$\\
IRASF15307+3252&$48.28^{+12.51}_{-13.43}$&$91.14^{+47.50}_{-42.15}$&$13.22^{+5.01}_{-5.36}$&$2.69^{+0.88}_{-1.12}$&$94.57^{+48.90}_{-43.55}$&$35.07^{+23.06}_{-17.97}$&$46.14^{+0.27}_{-0.43}$\\
\noalign{\smallskip}\hline\noalign{\smallskip}
\end{tabular}

  \label{tab:AGN}
\end{table*}

\begin{figure*}
  \centering 
  \subfigure[]
  {
  \includegraphics[scale =0.7] {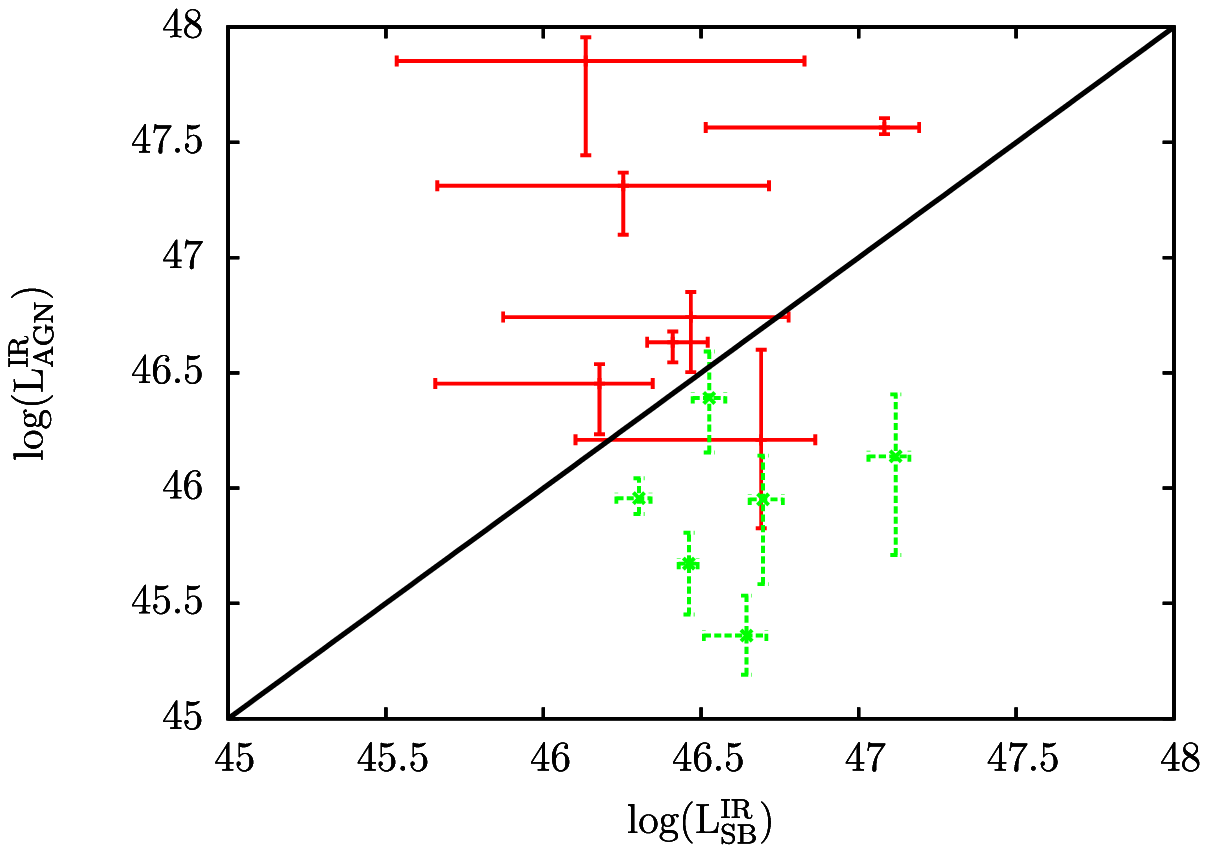}
  \label{fig:relation1}
  }
  \qquad
  \subfigure[]
  {
  \includegraphics[scale =0.7] {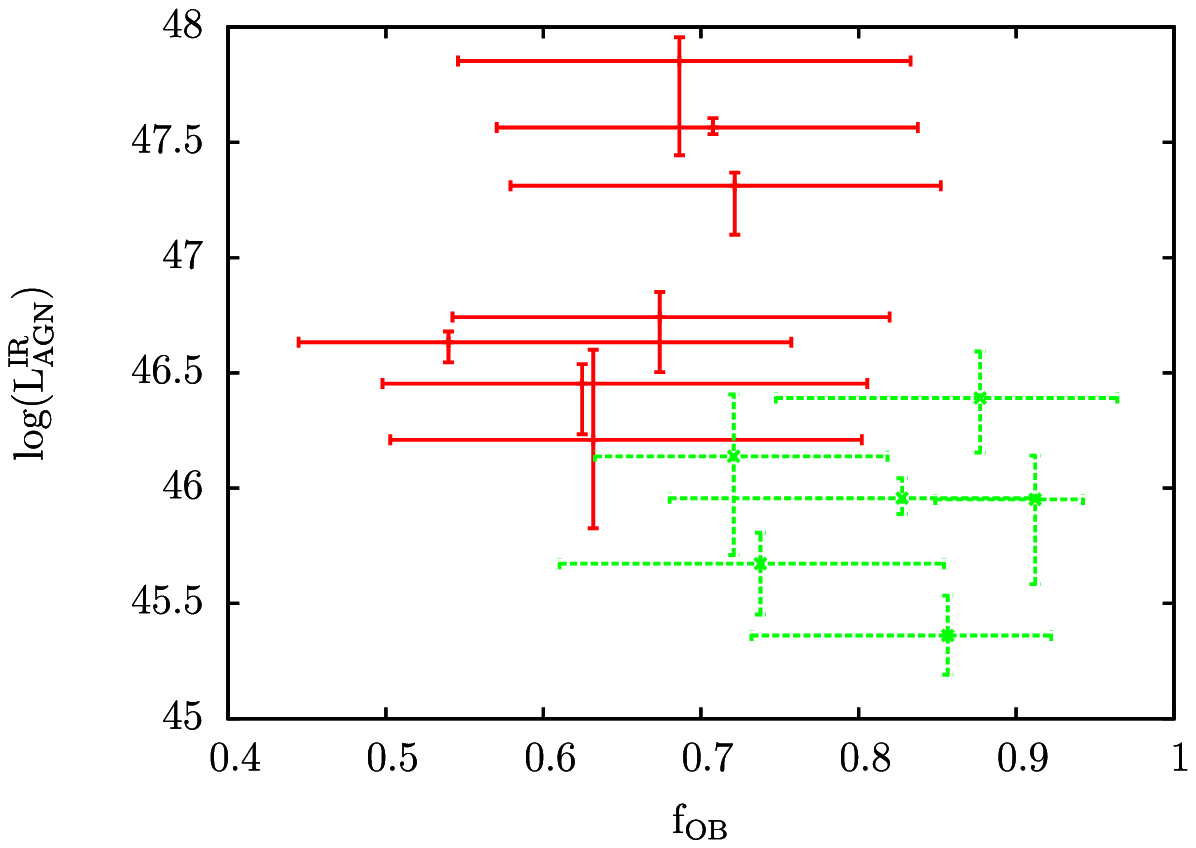}
  \label{fig:relation2}
  }
  \\
  \subfigure[]
  {
  \includegraphics[scale =0.7] {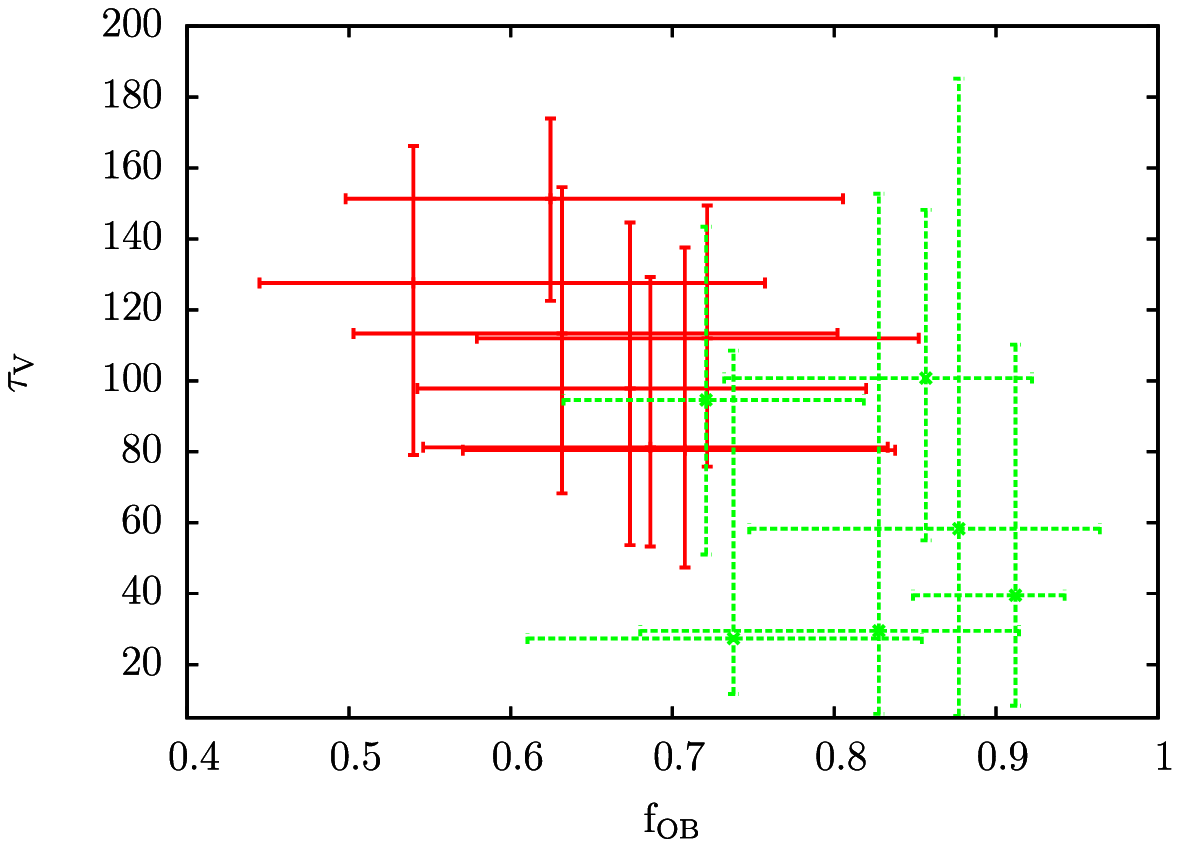}
  \label{fig:relation3}
  }
  \qquad
  \subfigure[]
  {
  \includegraphics[scale =0.7] {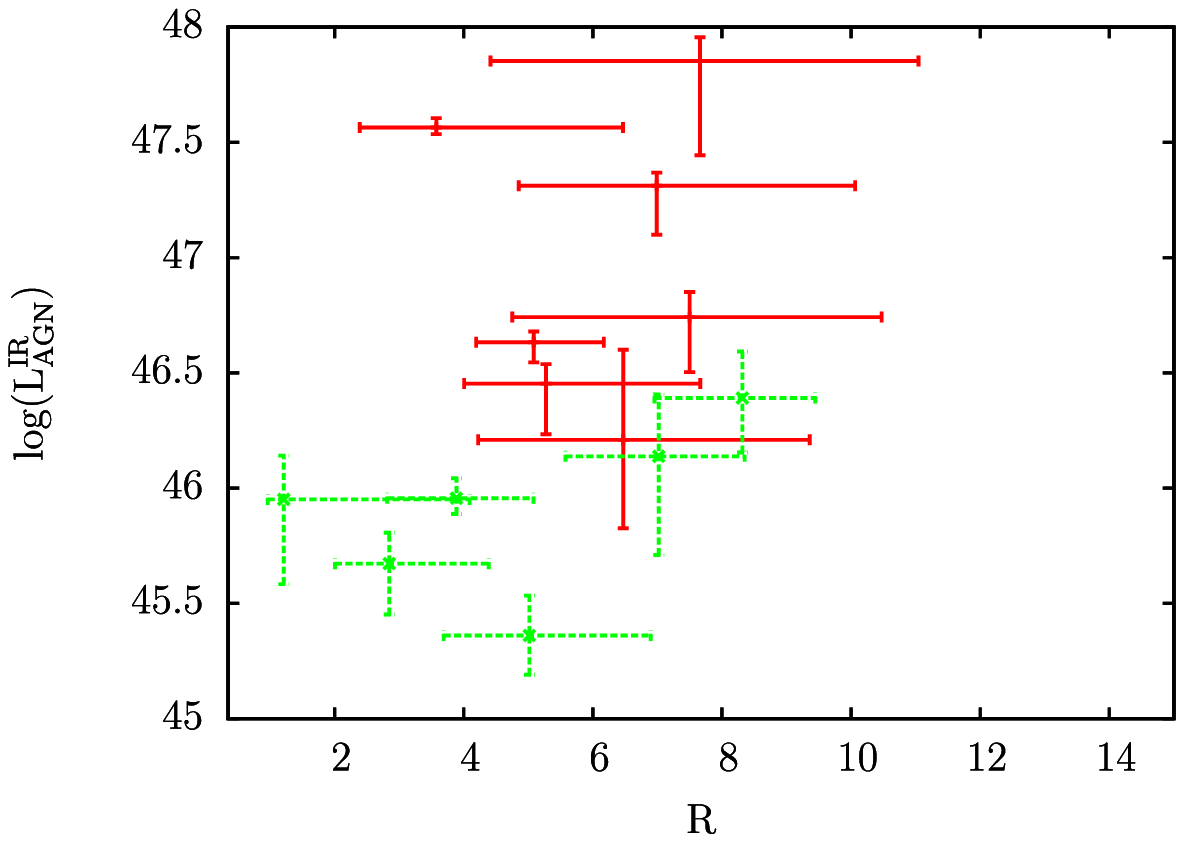}
  \label{fig:relation4}
  }
  \caption{Some relations between starburst and AGN parameters.
  The red-solid and green-dashed error bars are results for class A and B HLIRG, respectively.
  (a) The IR luminosity of starburst and AGN for all HLIRGs in the sample.
  The black-solid line represents position where the IR luminosity of starburst and AGN are equal.
  (b) The relation between the IR luminosity of AGN and the fraction of OB star in the starburst region.
  (c) The relation between the optical depth of clumps in AGN torus and the fraction of OB star in the starburst region.
  (d) The relation between the IR luminosity of AGN and the size of starburst region.
}
  \label{fig:relations}
\end{figure*}

In Table~\ref{tab:SB} and \ref{tab:AGN}, we present the estimated starburst and AGN parameters for class A and B HLIRGs by employing the `SB+AGN' model.
With the estimated starburst and AGN parameters of these HLIRGs, it would be interesting to explore some possible relations between these parameters, especially those between starburst and AGN.
However, with the very limited observations not all parameters can be well constrained.
In Figure \ref{fig:relations}, we present some relations between the starburst and AGN parameters that are relatively better constrained.

Figure \ref{fig:relation1} show the IR luminosity of starburst and AGN for all HLIRGs in the sample.
As shown clearly, the IR luminosity of most class A HLIRGs are dominated by AGN, while the IR luminosity of class B HLIRGs are dominated by starburst.
This is consistent with the conclusions drawn according to the Bayesian evidences as shown in Section~\ref{sssect:evidences}.
\cite{Ruiz2010a} classified the HLIRGs in their sample into class A and B according to the shape of their SEDs.
So, our results show that the class A and B HLIRGs essentially differ in their dominating emission source.

Figure \ref{fig:relation2} show the relation between the IR luminosity of AGN and the fraction of OB star in the starburst region.
The figure shows an anti-correlation between the fraction of OB star in the starburst region and the IR luminosity of AGN in the center.
This implies that the starburst in class B HLIRG is younger than that in class A HLIRG.
On the other hand, Figure \ref{fig:relation3} also show an anti-correlation between the optical depth of clumps in AGN torus and the fraction of OB star in the starburst region.
This may imply that the AGN torus in class A HLIRGs are more dusty than those in class B HLIRGs.
Furthermore, the results in Figure \ref{fig:relation4} show that the starburst region in class B HLIRG seems more compact than that in class A HLIRG.

\section{SUMMARY}
\label{sect:summary}
Dust-obscured starburst-AGN composite galaxies, such as ULIRGs and HLIRGs, represent important phases in the formation and evolution of galaxies.
It is still very challenging to understand the nature of these interesting but complex galaxies from their SEDs.
This can be achieved from the interplay between modeling and fitting of their SEDs.
However, a self-consistent multi-wavelength SED model for such complex systems must contain many parameters, and can only be established step by step.
So, a flexible, efficient and robust SED fitting tool is  very necessary.
In light of these, we developed a suite of general purpose programs, called BayeSED, for doing Bayesian analysis of SEDs. 
The PCA and ANN techniques are employed to allow an accurate and efficient generation of model SEDs.
Meanwhile, the state-of-art Bayesian inference tool, MultiNest, is interfaced with ANN to allow a highly efficient sampling of posterior distributions and the calculation of Bayesian evidence.

As a demonstration, we apply this code to a HLIRG sample.
By employing three models, we present a complete Bayesian analysis of their SEDs, including model comparison and parameter estimation.
According to the computed Bayesian evidence of different models and the estimated IR luminosity of starburst and AGN, we found that the class A and B HLIRG as defined by \cite{Ruiz2010a} essentially differ in their dominating emission source.
On the other hand, we found some relations between the estimated starburst and AGN parameters.
For example, the AGN torus of the HLIRGs dominated by AGN tend to be more dusty than that of HLIRGs dominated by starburst.
The starburst region of the HLIRGs dominated by starburst tends to be more compact and has a higher fraction of OB star than that of HLIRGs dominated by AGN.

These results are understandable in the context of galaxy merger driving starburst and delayed AGN activity \citep{Genzel1998a,Sanders1988a,Kauffmann2000a,DiMatteo2005a,Springel2005a,Hopkins2006b,Hopkins2008a}.
There may be an evolution path from class B HLIRG to class A HLIRG.
The class B HLIRG may represent the stage where a powerful AGN  has already been triggered but still not outshine the starburst, while in the state represented by class A HLIRG, the powerful AGN in the center becomes dominating the output of energy.
However, the sample studied here is still very small.
Further studies based on more complete samples of HLIRGs and more theoretical models are needed to verify this hypothesis.

Generally, we believe BayeSED can be a useful tool for understanding the nature of complex systems, such as dust-obscured starburst-AGN composite galaxies, from decoding their SEDs.
In the future works, we will apply this code to other larger samples to explore the interplay between starburst and AGN activities in these interesting galaxies.
On the other hand, there is still no well established SED models specifically for starburst-AGN composite galaxies.
So, it would be interesting to explore if a self-consistent SED model specifically for composite galaxies can have higher Bayesian evidence than a linear combination of Starburst+AGN models.

\acknowledgments
We thank an anonymous referee for his/her valuable comments which help to improve the paper greatly.
We thank Asensio Ramos, A. for sending their PCA and ANN routines, which are nice references for our work.
This work is supported by the National Natural Science Foundation of China (Grant Nos. 11033008 and 11103072), and the Chinese Academy of Sciences (Grant No. KJCX2-YW-T24).

\bibliography{ms.bbl}

\begin{thebibliography}{72}
\expandafter\ifx\csname natexlab\endcsname\relax\def\natexlab#1{#1}\fi

\bibitem[{{Almeida} {et~al.}(2010){Almeida}, {Baugh}, {Lacey}, {Frenk},
  {Granato}, {Silva}, \& {Bressan}}]{Almeida2010a}
{Almeida}, C., {Baugh}, C.~M., {Lacey}, C.~G., {Frenk}, C.~S., {Granato},
  G.~L., {Silva}, L., \& {Bressan}, A. 2010, \mnras, 402, 544

\bibitem[{{Andreon} {et~al.}(2000){Andreon}, {Gargiulo}, {Longo},
  {Tagliaferri}, \& {Capuano}}]{Andreon2000a}
{Andreon}, S., {Gargiulo}, G., {Longo}, G., {Tagliaferri}, R., \& {Capuano}, N.
  2000, \mnras, 319, 700

\bibitem[{{Asensio Ramos} \& {Ramos Almeida}(2009)}]{AsensioRamos2009a}
{Asensio Ramos}, A., \& {Ramos Almeida}, C. 2009, \apj, 696, 2075

\bibitem[{{Auld} {et~al.}(2008){Auld}, {Bridges}, \& {Hobson}}]{Auld2008a}
{Auld}, T., {Bridges}, M., \& {Hobson}, M.~P. 2008, \mnras, 387, 1575

\bibitem[{{Baes} {et~al.}(2003){Baes}, {Davies}, {Dejonghe}, {Sabatini},
  {Roberts}, {Evans}, {Linder}, {Smith}, \& {de Blok}}]{Baes2003a}
{Baes}, M., {et~al.} 2003, \mnras, 343, 1081

\bibitem[{{Bailer-Jones}(2011)}]{Bailer-Jones2011a}
{Bailer-Jones}, C.~A.~L. 2011, \mnras, 411, 435

\bibitem[{{Ben{\'{\i}}tez}(2000)}]{Benitez2000a}
{Ben{\'{\i}}tez}, N. 2000, \apj, 536, 571

\bibitem[{{Bertin} \& {Arnouts}(1996)}]{Bertin1996a}
{Bertin}, E., \& {Arnouts}, S. 1996, \aaps, 117, 393

\bibitem[{{Bruzual} \& {Charlot}(2003)}]{Bruzual2003a}
{Bruzual}, G., \& {Charlot}, S. 2003, \mnras, 344, 1000

\bibitem[{{Bruzual A.}(1983)}]{Bruzual1983a}
{Bruzual A.}, G. 1983, \apj, 273, 105

\bibitem[{{Budav{\'a}ri} {et~al.}(2009){Budav{\'a}ri}, {Wild}, {Szalay},
  {Dobos}, \& {Yip}}]{Budavari2009a}
{Budav{\'a}ri}, T., {Wild}, V., {Szalay}, A.~S., {Dobos}, L., \& {Yip}, C.-W.
  2009, \mnras, 394, 1496

\bibitem[{{Carballo} {et~al.}(2008){Carballo}, {Gonz{\'a}lez-Serrano}, {Benn},
  \& {Jim{\'e}nez-Luj{\'a}n}}]{Carballo2008a}
{Carballo}, R., {Gonz{\'a}lez-Serrano}, J.~I., {Benn}, C.~R., \&
  {Jim{\'e}nez-Luj{\'a}n}, F. 2008, \mnras, 391, 369

\bibitem[{{Chakrabarti} \& {Whitney}(2009)}]{Chakrabarti2009a}
{Chakrabarti}, S., \& {Whitney}, B.~A. 2009, \apj, 690, 1432

\bibitem[{{Collister} \& {Lahav}(2004)}]{Collister2004a}
{Collister}, A.~A., \& {Lahav}, O. 2004, \pasp, 116, 345

\bibitem[{{Conroy} {et~al.}(2009){Conroy}, {Gunn}, \& {White}}]{Conroy2009a}
{Conroy}, C., {Gunn}, J.~E., \& {White}, M. 2009, \apj, 699, 486

\bibitem[{Cybenko(1989)}]{Cybenko1989a}
Cybenko, G. 1989, Mathematics of Control, Signals, and Systems (MCSS), 2, 303,
  10.1007/BF02551274

\bibitem[{{da Cunha} {et~al.}(2008){da Cunha}, {Charlot}, \&
  {Elbaz}}]{daCunha2008a}
{da Cunha}, E., {Charlot}, S., \& {Elbaz}, D. 2008, \mnras, 388, 1595

\bibitem[{{Devriendt} {et~al.}(1999){Devriendt}, {Guiderdoni}, \&
  {Sadat}}]{Devriendt1999a}
{Devriendt}, J.~E.~G., {Guiderdoni}, B., \& {Sadat}, R. 1999, \aap, 350, 381

\bibitem[{{Di Matteo} {et~al.}(2005){Di Matteo}, {Springel}, \&
  {Hernquist}}]{DiMatteo2005a}
{Di Matteo}, T., {Springel}, V., \& {Hernquist}, L. 2005, \nat, 433, 604

\bibitem[{{Dullemond} \& {Dominik}(2004)}]{Dullemond2004a}
{Dullemond}, C.~P., \& {Dominik}, C. 2004, \aap, 417, 159

\bibitem[{{Efstathiou} {et~al.}(2000){Efstathiou}, {Rowan-Robinson}, \&
  {Siebenmorgen}}]{Efstathiou2000a}
{Efstathiou}, A., {Rowan-Robinson}, M., \& {Siebenmorgen}, R. 2000, \mnras,
  313, 734

\bibitem[{{Feldmann} {et~al.}(2006){Feldmann}, {Carollo}, {Porciani}, {Lilly},
  {Capak}, {Taniguchi}, {Le F{\`e}vre}, {Renzini}, {Scoville}, {Ajiki},
  {Aussel}, {Contini}, {McCracken}, {Mobasher}, {Murayama}, {Sanders},
  {Sasaki}, {Scarlata}, {Scodeggio}, {Shioya}, {Silverman}, {Takahashi},
  {Thompson}, \& {Zamorani}}]{Feldmann2006a}
{Feldmann}, R., {et~al.} 2006, \mnras, 372, 565

\bibitem[{{Feroz} \& {Hobson}(2008)}]{Feroz2008a}
{Feroz}, F., \& {Hobson}, M.~P. 2008, \mnras, 384, 449

\bibitem[{{Feroz} {et~al.}(2009){Feroz}, {Hobson}, \& {Bridges}}]{Feroz2009a}
{Feroz}, F., {Hobson}, M.~P., \& {Bridges}, M. 2009, \mnras, 398, 1601

\bibitem[{{Fioc} \& {Rocca-Volmerange}(1997)}]{Fioc1997a}
{Fioc}, M., \& {Rocca-Volmerange}, B. 1997, \aap, 326, 950

\bibitem[{{Firth} {et~al.}(2003){Firth}, {Lahav}, \& {Somerville}}]{Firth2003a}
{Firth}, A.~E., {Lahav}, O., \& {Somerville}, R.~S. 2003, \mnras, 339, 1195

\bibitem[{{Francis} {et~al.}(1992){Francis}, {Hewett}, {Foltz}, \&
  {Chaffee}}]{Francis1992a}
{Francis}, P.~J., {Hewett}, P.~C., {Foltz}, C.~B., \& {Chaffee}, F.~H. 1992,
  \apj, 398, 476

\bibitem[{{Genzel} {et~al.}(1998){Genzel}, {Lutz}, {Sturm}, {Egami}, {Kunze},
  {Moorwood}, {Rigopoulou}, {Spoon}, {Sternberg}, {Tacconi-Garman}, {Tacconi},
  \& {Thatte}}]{Genzel1998a}
{Genzel}, R., {et~al.} 1998, \apj, 498, 579

\bibitem[{{Glazebrook} {et~al.}(1998){Glazebrook}, {Offer}, \&
  {Deeley}}]{Glazebrook1998a}
{Glazebrook}, K., {Offer}, A.~R., \& {Deeley}, K. 1998, \apj, 492, 98

\bibitem[{{Granato} {et~al.}(2000){Granato}, {Lacey}, {Silva}, {Bressan},
  {Baugh}, {Cole}, \& {Frenk}}]{Granato2000a}
{Granato}, G.~L., {Lacey}, C.~G., {Silva}, L., {Bressan}, A., {Baugh}, C.~M.,
  {Cole}, S., \& {Frenk}, C.~S. 2000, \apj, 542, 710

\bibitem[{{Groves} {et~al.}(2008){Groves}, {Dopita}, {Sutherland}, {Kewley},
  {Fischera}, {Leitherer}, {Brandl}, \& {van Breugel}}]{Groves2008a}
{Groves}, B., {Dopita}, M.~A., {Sutherland}, R.~S., {Kewley}, L.~J.,
  {Fischera}, J., {Leitherer}, C., {Brandl}, B., \& {van Breugel}, W. 2008,
  \apjs, 176, 438

\bibitem[{{Han} {et~al.}(2007){Han}, {Podsiadlowski}, \&
  {Lynas-Gray}}]{Han2007a}
{Han}, Z., {Podsiadlowski}, P., \& {Lynas-Gray}, A.~E. 2007, \mnras, 380, 1098

\bibitem[{Haykin(1999)}]{Haykin1999a}
Haykin, S. 1999, {N}eural {N}etworks: {A} {C}omprehensive {F}oundation (Upper
  Saddle River, NJ: Prentice Hall), 2nd edition

\bibitem[{{Hopkins} {et~al.}(2006){Hopkins}, {Hernquist}, {Cox}, {Di Matteo},
  {Robertson}, \& {Springel}}]{Hopkins2006b}
{Hopkins}, P.~F., {Hernquist}, L., {Cox}, T.~J., {Di Matteo}, T., {Robertson},
  B., \& {Springel}, V. 2006, \apjs, 163, 1

\bibitem[{{Hopkins} {et~al.}(2008){Hopkins}, {Hernquist}, {Cox}, \& {Kere{\v
  s}}}]{Hopkins2008a}
{Hopkins}, P.~F., {Hernquist}, L., {Cox}, T.~J., \& {Kere{\v s}}, D. 2008,
  \apjs, 175, 356

\bibitem[{{Jonsson}(2006)}]{Jonsson2006a}
{Jonsson}, P. 2006, \mnras, 372, 2

\bibitem[{{Kauffmann} \& {Haehnelt}(2000)}]{Kauffmann2000a}
{Kauffmann}, G., \& {Haehnelt}, M. 2000, \mnras, 311, 576

\bibitem[{{Kauffmann} {et~al.}(2003){Kauffmann}, {Heckman}, {White}, {Charlot},
  {Tremonti}, {Brinchmann}, {Bruzual}, {Peng}, {Seibert}, {Bernardi},
  {Blanton}, {Brinkmann}, {Castander}, {Cs{\'a}bai}, {Fukugita}, {Ivezic},
  {Munn}, {Nichol}, {Padmanabhan}, {Thakar}, {Weinberg}, \&
  {York}}]{Kauffmann2003b}
{Kauffmann}, G., {et~al.} 2003, \mnras, 341, 33

\bibitem[{Kurt \& Hornik(1991)}]{Kurt1991a}
Kurt, \& Hornik. 1991, Neural Networks, 4, 251

\bibitem[{{Lacey} \& {Silk}(1991)}]{Lacey1991a}
{Lacey}, C., \& {Silk}, J. 1991, \apj, 381, 14

\bibitem[{{Lahav} {et~al.}(1996){Lahav}, {Naim}, {Sodr{\'e}}, \&
  {Storrie-Lombardi}}]{Lahav1996a}
{Lahav}, O., {Naim}, A., {Sodr{\'e}}, Jr., L., \& {Storrie-Lombardi}, M.~C.
  1996, \mnras, 283, 207

\bibitem[{{Larson} \& {Tinsley}(1978)}]{Larson1978a}
{Larson}, R.~B., \& {Tinsley}, B.~M. 1978, \apj, 219, 46

\bibitem[{{Leitherer} {et~al.}(1999){Leitherer}, {Schaerer}, {Goldader},
  {Gonz{\'a}lez Delgado}, {Robert}, {Kune}, {de Mello}, {Devost}, \&
  {Heckman}}]{Leitherer1999a}
{Leitherer}, C., {et~al.} 1999, \apjs, 123, 3

\bibitem[{{Maraston}(2005)}]{Maraston2005a}
{Maraston}, C. 2005, \mnras, 362, 799

\bibitem[{{Mukherjee} {et~al.}(2006){Mukherjee}, {Parkinson}, \&
  {Liddle}}]{Mukherjee2006a}
{Mukherjee}, P., {Parkinson}, D., \& {Liddle}, A.~R. 2006, \apjl, 638, L51

\bibitem[{{Nenkova} {et~al.}(2008{\natexlab{a}}){Nenkova}, {Sirocky},
  {Ivezi{\'c}}, \& {Elitzur}}]{Nenkova2008a}
{Nenkova}, M., {Sirocky}, M.~M., {Ivezi{\'c}}, {\v Z}., \& {Elitzur}, M.
  2008{\natexlab{a}}, \apj, 685, 147

\bibitem[{{Nenkova} {et~al.}(2008{\natexlab{b}}){Nenkova}, {Sirocky},
  {Nikutta}, {Ivezi{\'c}}, \& {Elitzur}}]{Nenkova2008b}
{Nenkova}, M., {Sirocky}, M.~M., {Nikutta}, R., {Ivezi{\'c}}, {\v Z}., \&
  {Elitzur}, M. 2008{\natexlab{b}}, \apj, 685, 160

\bibitem[{{Noll} {et~al.}(2009){Noll}, {Burgarella}, {Giovannoli}, {Buat},
  {Marcillac}, \& {Mu{\~n}oz-Mateos}}]{Noll2009a}
{Noll}, S., {Burgarella}, D., {Giovannoli}, E., {Buat}, V., {Marcillac}, D., \&
  {Mu{\~n}oz-Mateos}, J.~C. 2009, \aap, 507, 1793

\bibitem[{{Rowan-Robinson}(2000)}]{Rowan-Robinson2000a}
{Rowan-Robinson}, M. 2000, \mnras, 316, 885

\bibitem[{{Ruiz} {et~al.}(2007){Ruiz}, {Carrera}, \& {Panessa}}]{Ruiz2007a}
{Ruiz}, A., {Carrera}, F.~J., \& {Panessa}, F. 2007, \aap, 471, 775

\bibitem[{{Ruiz} {et~al.}(2010){Ruiz}, {Miniutti}, {Panessa}, \&
  {Carrera}}]{Ruiz2010a}
{Ruiz}, A., {Miniutti}, G., {Panessa}, F., \& {Carrera}, F.~J. 2010, \aap, 515,
  A99+

\bibitem[{{Salim} {et~al.}(2007){Salim}, {Rich}, {Charlot}, {Brinchmann},
  {Johnson}, {Schiminovich}, {Seibert}, {Mallery}, {Heckman}, {Forster},
  {Friedman}, {Martin}, {Morrissey}, {Neff}, {Small}, {Wyder}, {Bianchi},
  {Donas}, {Lee}, {Madore}, {Milliard}, {Szalay}, {Welsh}, \&
  {Yi}}]{Salim2007a}
{Salim}, S., {et~al.} 2007, \apjs, 173, 267

\bibitem[{{Sanders} {et~al.}(1988){Sanders}, {Soifer}, {Elias}, {Madore},
  {Matthews}, {Neugebauer}, \& {Scoville}}]{Sanders1988a}
{Sanders}, D.~B., {Soifer}, B.~T., {Elias}, J.~H., {Madore}, B.~F., {Matthews},
  K., {Neugebauer}, G., \& {Scoville}, N.~Z. 1988, \apj, 325, 74

\bibitem[{{Searle} {et~al.}(1973){Searle}, {Sargent}, \&
  {Bagnuolo}}]{Searle1973a}
{Searle}, L., {Sargent}, W.~L.~W., \& {Bagnuolo}, W.~G. 1973, \apj, 179, 427

\bibitem[{{Shaw} {et~al.}(2007){Shaw}, {Bridges}, \& {Hobson}}]{Shaw2007a}
{Shaw}, J.~R., {Bridges}, M., \& {Hobson}, M.~P. 2007, \mnras, 378, 1365

\bibitem[{{Siebenmorgen} \& {Kr{\"u}gel}(2007)}]{Siebenmorgen2007a}
{Siebenmorgen}, R., \& {Kr{\"u}gel}, E. 2007, \aap, 461, 445

\bibitem[{{Silva} {et~al.}(1998){Silva}, {Granato}, {Bressan}, \&
  {Danese}}]{Silva1998a}
{Silva}, L., {Granato}, G.~L., {Bressan}, A., \& {Danese}, L. 1998, \apj, 509,
  103

\bibitem[{{Silva} {et~al.}(2011){Silva}, {Schurer}, {Granato}, {Almeida},
  {Baugh}, {Frenk}, {Lacey}, {Paoletti}, {Petrella}, \&
  {Selvestrel}}]{Silva2011a}
{Silva}, L., {et~al.} 2011, \mnras, 410, 2043

\bibitem[{{Skilling}(2004)}]{Skilling2004a}
{Skilling}, J. 2004, in American Institute of Physics Conference Series, Vol.
  735, American Institute of Physics Conference Series, ed. {R.~Fischer,
  R.~Preuss, \& U.~V.~Toussaint}, 395--405

\bibitem[{{Springel} {et~al.}(2005){Springel}, {Di Matteo}, \&
  {Hernquist}}]{Springel2005a}
{Springel}, V., {Di Matteo}, T., \& {Hernquist}, L. 2005, \mnras, 361, 776

\bibitem[{{Tinsley}(1972)}]{Tinsley1972a}
{Tinsley}, B.~M. 1972, \aap, 20, 383

\bibitem[{{Tinsley}(1978)}]{Tinsley1978a}
---. 1978, \apj, 222, 14

\bibitem[{{Trotta}(2008)}]{Trotta2008a}
{Trotta}, R. 2008, Contemporary Physics, 49, 71

\bibitem[{{Tuffs} {et~al.}(2004){Tuffs}, {Popescu}, {V{\"o}lk}, {Kylafis}, \&
  {Dopita}}]{Tuffs2004a}
{Tuffs}, R.~J., {Popescu}, C.~C., {V{\"o}lk}, H.~J., {Kylafis}, N.~D., \&
  {Dopita}, M.~A. 2004, \aap, 419, 821

\bibitem[{{Vanzella} {et~al.}(2004){Vanzella}, {Cristiani}, {Fontana},
  {Nonino}, {Arnouts}, {Giallongo}, {Grazian}, {Fasano}, {Popesso}, {Saracco},
  \& {Zaggia}}]{Vanzella2004a}
{Vanzella}, E., {et~al.} 2004, \aap, 423, 761

\bibitem[{{Walcher} {et~al.}(2011){Walcher}, {Groves}, {Budav{\'a}ri}, \&
  {Dale}}]{Walcher2011a}
{Walcher}, J., {Groves}, B., {Budav{\'a}ri}, T., \& {Dale}, D. 2011, \apss,
  331, 1

\bibitem[{{White} \& {Frenk}(1991)}]{White1991a}
{White}, S.~D.~M., \& {Frenk}, C.~S. 1991, \apj, 379, 52

\bibitem[{{White} \& {Rees}(1978)}]{White1978a}
{White}, S.~D.~M., \& {Rees}, M.~J. 1978, \mnras, 183, 341

\bibitem[{{Wild} \& {Hewett}(2005)}]{Wild2005a}
{Wild}, V., \& {Hewett}, P.~C. 2005, \mnras, 358, 1083

\bibitem[{{Wild} {et~al.}(2007){Wild}, {Kauffmann}, {Heckman}, {Charlot},
  {Lemson}, {Brinchmann}, {Reichard}, \& {Pasquali}}]{Wild2007a}
{Wild}, V., {Kauffmann}, G., {Heckman}, T., {Charlot}, S., {Lemson}, G.,
  {Brinchmann}, J., {Reichard}, T., \& {Pasquali}, A. 2007, \mnras, 381, 543

\bibitem[{{Zhang} {et~al.}(2005{\natexlab{a}}){Zhang}, {Han}, {Li}, \&
  {Hurley}}]{Zhang2005a}
{Zhang}, F., {Han}, Z., {Li}, L., \& {Hurley}, J.~R. 2005{\natexlab{a}},
  \mnras, 357, 1088

\bibitem[{{Zhang} {et~al.}(2005{\natexlab{b}}){Zhang}, {Li}, \&
  {Han}}]{Zhang2005b}
{Zhang}, F., {Li}, L., \& {Han}, Z. 2005{\natexlab{b}}, \mnras, 364, 503

\end{thebibliography}

\clearpage

\end{document}